\documentclass[]{article}
\usepackage{epsfig}
\newcommand{\bm}{\bibitem}
\newcommand{\cp}{\chi^{(+)}}
\newcommand{\cm}{\chi^{(-)*}}

\newcommand{\vv}{V_{bc}({\bf r}_1)}
\newcommand{\ri}{{\bf r}_i}
\newcommand{\ro}{{\bf r}_1}
\newcommand{\ak}{{\bf k}_a}
\newcommand{\bq}{{\bf k}_b}

\newcommand{\rc}{{\bf r}_c}
\newcommand{\cq}{{\bf k}_c}
\newcommand{\we}{\Psi^{(+)}_a(\xi_a,{\bf r}_1,{\bf r}_i)}
\newcommand{\fa}{ _2F_1(1-i\eta_a,1-i\eta_b;2;D(0))} 
\newcommand{\fB}{ _2F_1(-i\eta_a,-i\eta_b;1;D(0))}

\begin{document}

%\draft
%\tighten
\begin{center}
\Huge{\bf BREAKUP REACTIONS OF DRIP LINE NUCLEI{$^1$}} \\
%\date{\today}
\vskip2cm
\Large{\bf R. Shyam and R. Chatterjee} \\
\vskip1cm
{\it {
Saha Institute of Nuclear Physics, 1/AF Bidhan Nagar, Kolkata 700064,
India} }\\
\vskip2cm 
{{$^1$}Lectures presented in the workshop on} \\
\bf ``Nuclei at Extremes of Isospin and Mass'', \\
\bf held at Toshali Sands, Puri, \\
\bf March 11 -- 22, 2003.
%\maketitle
\end{center}
\tableofcontents
\newpage
\begin{center}
{\bf \large ABSTRACT}
\end{center}

\noindent
The formal theory of breakup reactions is reviewed.
The direct 
breakup mechanism which is formulated within the
framework of the post form distorted wave Born approximation
is discussed in detail. In this theory, which requires the
information about only the ground state wave function of the projectile,
the fragment-target interactions 
are included to all orders while fragment-fragment interaction is treated only in the first order.  
The general applicability of this theory to
describe the breakup of halo nuclei is demonstrated by comparing the
calculations with data for total, as well as energy and angle
integrated cross sections and momentum distributions
of fragments in reactions induced by a number of
halo nuclei.
We investigate
the role played by the pure Coulomb, pure nuclear and the Coulomb-nuclear
interference terms. 
Postacceleration effects in the Coulomb breakup of neutron halo 
nuclei is also studied.

\newpage
%%%%%%%%%%%%%%%%%% B1 %%%%%%%%%%%%%%%%%%%%%%%%%%%%%%%%%%%%%%%%%%%%%%%5
\section{Introduction}

Recently advances made in accelerator and related technologies have
provided us the opportunity to produce and work with nuclei 
having very short half-lives and very small one- or two-nucleon
separation energies [1-14]. These nuclei lie very close to drip lines
(the limit of neutron or proton binding). Nuclei at extremes of binding
can exhibit behaviours which are quite different from those of the stable
isotopes. We still lack a fully microscopic understanding of the stability
of these unique many body systems. These nuclei are
important also from the nuclear astrophysics point of view. The
rapid neutron capture (the r-process) together with the slow neutron
capture (the s-process), which are dominant mechanisms for the nucleosynthesis
of heavy elements above iron pass mostly through the
neutron rich region. The properties of these nuclei are important inputs
to theoretical calculations on stellar burning which otherwise are often
forced to rely on global assumptions about nuclear masses, decays and
level structures extracted from stable nuclei.

The first round of measurements involving neutron rich nuclei
{\cite{tani85,mittig,st-lau,fukuda91,gil01,ritu02}} have confirmed the 
existence of a novel structure in some of them in which 
a low density tail of loosely bound neutrons extends too far out in
the coordinate space as compared to a stable core
(also known as the neutron halo{\footnote {The
term `neutron halo' was introduced by Burhop et al. {\cite{bur}},
in the context (now called `neutron skin') of the bulk of the neutron
density extending further out in space than the proton density.}).
The quantum mechanical tunneling of very loosely bound 
valence neutrons leads to the formation of such a structure. 
The existence of neutron halo has been confirmed in 
$^{11}$Be {\cite{ex11be}}, $^{14}$B {\cite{bazz98,ex14b}}, 
$^{19}$C {\cite{ex19c}} (one-neutron halo), and
$^{6}$He, $^{11}$Li {\cite{tani85,ex2,ex3}},  
$^{14}$Be {\cite{ex4,ex5}}, and $^{17}$B {\cite{ex4}} (two-neutron halo).
Recently, some proton halo nuclei have also been identified
($^{8}$B {\cite{ex6,ex7,ex8}}, $^{17}$Ne {\cite{ex9}}, $^{20}$Mg {\cite{ex10}},
and $^{26,27,28}$P {\cite{ex11}).

Halo nuclei, in most cases, have only one bound state (the ground state)
and a broad featureless continuum. Thus, methods of conventional nuclear 
structure studies, namely, measurements of energies and 
spin-parities of excited states are not applicable in these cases. 
However, due to their small binding energies, they can be
easily excited above their particle emission thresholds. Hence their 
breakup reactions in the Coulomb and nuclear fields of the 
target nuclei could be useful tools to investigate their structures.

To be able to extract reliable structure information of halo nuclei 
from the breakup data, it is quite desirable to have a 
theory of these reactions which (1) is fully quantum mechanical; (2) treats
the Coulomb and nuclear breakups as well as their interference terms 
consistently on an equal footing; (3) includes the recoil of the core
within the halo nucleus, and the finite range of the core-halo interaction,
and (4) involves least adjustable parameters.

We discuss the formal theory
for halo breakup reactions in section 2. We specially mention a theory 
formulated within the framework of the post form distorted wave
Born approximation (DWBA) where both Coulomb and nuclear breakups
can be treated consistently on an equal footing. The full ground
state wave function of the projectile enters as an input into
this theory. Thus, information about the halo structure can be 
extracted by comparing the calculations with the available data. This
is shown in section 3. We also discuss the role of nuclear breakup
and of Coulomb-nuclear interference (CNI) terms in this section. The post
form DWBA theory is uniquely suited to study the postacceleration 
effects in the halo breakup reactions. This is an higher order effect
which is studied in section 4. We give conclusions and the future outlook 
in section 5. 
%%%%%%%%%%%%%%%%%%%% E1 %%%%%%%%%%%%%%%%%%%%%%%%%%%%%%%%%%%%%%%%%
%%%%%%%%%%%%%%%%%%%% B2 %%%%%%%%%%%%%%%%%%%%%%%%%%%%%%%%%%%%%%%%%
\section{Formal Theory of Breakup Reactions}

%%%%%%%%%%%%%%%%%%%%%%%%%%%%%%%%%%%%%
%\section{A General introduction}
%%%%%%%%%%%%%%%%%%%%%%%%%%%%%%%%%%%%%

\begin{figure}[ht]
\begin{center}
\mbox{\epsfig{file=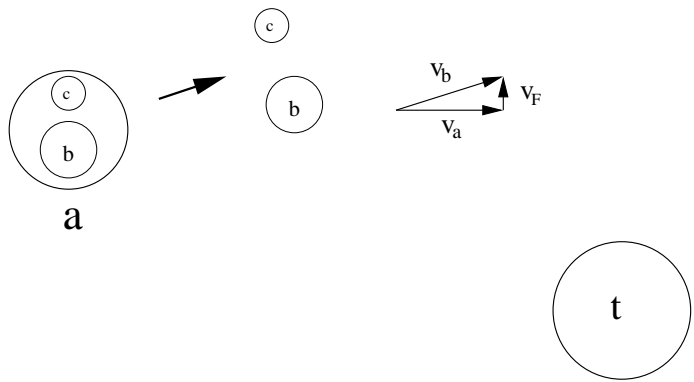,width=.75\textwidth}}
\end{center}
\caption{The direct breakup mechanism in a participant - spectator model. } 
\label{fig:c2_1)}
\end{figure}
The basic mechanism of the breakup reactions can be described in a simple
participant--spectator model in which the projectile $a$ which is supposed
to consist of two substructures, say, $b$ and $c$, interacts with 
a target $t$.
It might so happen that one of the substructures $b$ (the spectator) misses 
the target and keeps moving in its original direction while the 
substructure $c$ (the participant) interacts with it.
In such a situation the velocity of fragment $b$ (${\bf v}_b$) 
can be written as
\begin{eqnarray}
{\bf v}_b &=& {\bf v}_a + {\bf v}_F,  
\end{eqnarray}
where ${\bf v}_a$ is the velocity of projectile $a$, and ${\bf v}_F$
is the velocity associated with the Fermi motion of $b$ inside the
projectile before its breakup (see Fig. 1). Therefore
\begin{eqnarray}
v_b^2 &=& v_a^2 + v_F^2 + 2{\bf v}_a.{\bf v}_F~. 
\end{eqnarray}
As $v_F^2 << v_a^2$, we can neglect $v_F^2$ in Eq. (2) and get 
the maximum and minimum
values of $v_b^2$ (corresponding to $+$ and $-$ signs, respectively) to be 
given by 
\begin{eqnarray}
v_b^2 \approx v_a^2 \pm 2v_av_F. 
\end{eqnarray}
Hence, maximum and minimum possible energies of $b$ in the final channel are
given by
\begin{eqnarray}
E_b = {1\over 2}m_bv_b^2 = {m_b\over {m_a}}({1\over 2}m_av_a^2 \pm m_av_av_F),
\end{eqnarray}
where $m_b$ is the mass of fragment $b$, and $m_a$ is the mass of the 
projectile $a$. Defining $p_a = m_av_a$ and $p_F = m_bv_F$, we have
\begin{eqnarray}
E_b = {m_b\over {m_a}}E_a \pm {p_a p_F \over {m_a}}.
\end{eqnarray}
Thus, in the energy spectrum of
particle $b$ one would expect a peak at ${m_b\over {m_a}}E_a$, with a
width ($2{p_a p_F \over {m_a}}$) depending on its Fermi momentum 
inside the projectile.
This simple picture, which was proposed by Serber {\cite {serber}}
way back in 1947, is realized remarkably well in majority of the breakup data. 

Now, unless some kind of exclusive measurement is made in which
both fragments $b$ and $c$ are detected, the fragment $c$ can interact with
the target nucleus, $t$, in all possible ways. We distinguish between 
two classes of breakup processes: elastic breakup, where the target nucleus
remains in the ground state (the interaction $c-t$ is elastic) and inelastic
breakup where $c$ can interact in all possible ways with the target nucleus
(inelastic excitation
of the target, compound nucleus formation, transfer processes etc.). 

There is another picture of breakup
reactions in which the projectile is excited as a whole to a state in the
continuum which subsequently decays into fragments $b$ and $c$ as it
leaves the interaction zone. This process is called the
sequential breakup (or resonant breakup). Both
Coulomb and nuclear interactions between the projectile and target
can cause the inelastic excitation of the former.

In the next sub-section, we shall describe the transition amplitude and its various 
representations and introduce the distorted wave Born approximation. 
%%%%%%%%%%%%%%%%%%%%%%%%%%%%%%%%%%%%%%%%%%
\subsection{The Transition matrix and distorted wave Born approximation (DWBA)}
%%%%%%%%%%%%%%%%%%%%%%%%%%%%%%%%%%%%%%%%%%

Let us consider the reaction $ a + t \rightarrow b + c + t$, in a three
body model, where the projectile $a$, incident with momentum ${\bf q}_a$,
breaks up into fragments $b$ and $c$ with momenta ${\bf q}_b$ and ${\bf q}_c$, 
respectively in the Coulomb and nuclear fields of a target $t$. 

The Hamiltonian of the system is written as
\begin{eqnarray}
H = T_b + T_c + T_t + V_{bc} + V_{bt} + V_{ct},
\end{eqnarray}
where $T_i$ is the kinetic energy of particle $i$ and $V_{ij}$ is the
two-body interaction between $i$ and $j$; their separation will be denoted by
${\bf r}_{ij}$ in the following. 

To find the interaction in the initial and final channels, we note that the
asymptotic Hamiltonians in the initial (prior) and final (post) channels are 
\begin{eqnarray}
H_i=T_b + T_c + T_t + V_{bc}
\end{eqnarray}
 and 
\begin{eqnarray}
 H_f=T_b + T_c + T_t,
\end{eqnarray}
 respectively. 
Hence the initial (prior) channel interaction is 
\begin{eqnarray}
V_i=H-H_i=V_{bt}+V_{ct}
\end{eqnarray}
 and
the final (post) channel interaction is 
\begin{eqnarray}
V_f=H-H_f=V_{bc}+V_{bt}+V_{ct}.
\end{eqnarray}

There are two exact $T$ -- matrices {\cite{baur84}} 
\begin{eqnarray}
T^{(+)[post]}_{fi} = \langle e^{i{\bf q}_c.{\bf r}_{ct}} e^{i{\bf q}_b.{\bf r}_{bt}}| 
V_{bc}+V_{bt}+V_{ct} | \Psi_i^{(+)} \rangle, 
\end{eqnarray}
and
\begin{eqnarray}
T^{(-)[prior]}_{fi} = \langle \Psi_i^{(-)} | V_{bt}+V_{ct} | e^{i{\bf q}_a.{\bf r}_{at}}
\phi_{a}({\bf r}_{bc}) \rangle,
\end{eqnarray}
which are the starting points for a discussion on the theory of breakup
processes. The ground state wave function of the projectile, 
$\phi_{a}({\bf r}_{bc})$, satisfies 
\begin{eqnarray}
(T_b + T_c + V_{bc})\phi_{a}(r_{bc}) = -\epsilon_{a}\phi_{a}({\bf r}_{bc}),
\end{eqnarray}
where $\epsilon_{a}$ is the separation energy between fragments $b$ and $c$
in the ground state of the projectile. 
$\Psi_i^{(+)}$ is the exact three-body scattering 
wave function with {\it outgoing wave} boundary condition (denoted by (+) sign
in the superscript), and 
$\Psi_f^{(-)}$ is the exact three-body scattering 
wave function with {\it ingoing wave} boundary condition (denoted by (-) sign
in the superscript). They are the
exact eigenfunctions of the three-body Hamiltonian [Eq. (6)]. Thus they
satisfy 
\begin{eqnarray}
H\Psi_i^{(+)} = E\Psi_i^{(+)}
\end{eqnarray}
and
\begin{eqnarray}
H\Psi_f^{(-)} = E\Psi_f^{(-)},
\end{eqnarray}
where $E$ is the total energy of the system.
We now use the Gell-Mann -- Goldberger two potential formula {\cite{gell}} 
to rewrite Eqs. (11) and (12) as
\begin{eqnarray}
T^{(+)[post]}_{fi} = \langle \chi^{(-)}_{q_c}({\bf r}_{ct}) 
\chi^{(-)}_{q_b}({\bf r}_{bt})| 
V_{bc}+V_{bt}+V_{ct}-U_{bt}-U_{ct}| \Psi_i^{(+)} \rangle 
\end{eqnarray}
and
\begin{eqnarray}
T^{(-)[prior]}_{fi} = \langle \Psi_f^{(-)} |V_{ct} + V_{bt} -U_{at}| 
\chi^{(+)}_{q_a}({\bf r}_{at}) \phi_{a}({\bf r}_{bc})\rangle. 
\end{eqnarray}
In Eqs. (16) and (17), wave functions 
$\chi^{(-)}_{q_c}({\bf r}_{ct})\chi^{(-)}_{q_b}({\bf r}_{bt})$ and 
$\chi^{(+)}_{q_a}({\bf r}_{at})$ satisfy the Schr\"{o}dinger equations
\begin{eqnarray}
[T_{r_{ct}} + T_{r_{bt}} + U_{bt}+ U_{ct}]
\chi^{(-)}_{q_c}({\bf r}_{ct}) \chi^{(-)}_{q_b}({\bf r}_{bt}) =
E \chi^{(-)}_{q_c}({\bf r}_{ct}) \chi^{(-)}_{q_b}({\bf r}_{bt})
\end{eqnarray}
and
\begin{eqnarray}
[T_{r_{at}} + U_{at}]
\chi^{(+)}_{q_a}({\bf r}_{at}) = 
(E +\epsilon_{a}) \chi^{(+)}_{q_a}({\bf r}_{at}),
\end{eqnarray}
respectively. In Eqs. (18) and (19), $U_{it}$ are auxiliary potentials
acting between particle `$i$' and the target.

Assuming $V_{bt}=U_{bt}$ and $V_{ct}=U_{ct}$ in Eq. (16), 
we have {\cite{lev,vin}}
\begin{eqnarray}
T^{(+)[post]}_{fi} = \langle \chi^{(-)}_{q_c}({\bf r}_{ct}) 
\chi^{(-)}_{q_b}({\bf r}_{bt})| 
V_{bc}| \Psi_i^{(+)} \rangle. 
\end{eqnarray}

Let us now introduce the distorted wave Born 
Approximation (DWBA) {\cite{satchler}} for the exact wave functions 
$\Psi_i^{(+)}$ and $\Psi_f^{(-)}$ in Eqs. (20) and (17). 

If one assumes that the inelastic excitations 
of the projectile are small, then the wave function $\Psi_i^{(+)}$ can be
approximated by 
\begin{eqnarray}
\Psi_i^{(+)} \approx \chi^{(+)}_{q_a}({\bf r}_{at}) \phi_a({\bf r}_{bc}).
\end{eqnarray}
The post form DWBA $T$ -- matrix is then 
\begin{eqnarray}
T^{(+)[post]}_{fi}(DWBA) = 
\langle \chi^{(-)}_{q_b}({\bf r}_{bt}) \chi^{(-)}_{q_c}({\bf r}_{ct})
|V_{bc} | \chi^{(+)}_{q_a}({\bf r}_{at}) \phi_a({\bf r}_{bc})\rangle. 
\end{eqnarray}

If, on the other hand, one assumes that the final state interaction 
between the
breakup fragments ($b$ and $c$) is not important, i.e. $V_{bc}$ 
is weak in the final channel, then one can write the 
exact wave function $\Psi_f^{(-)}$ as,
\begin{eqnarray}
\Psi_f^{(-)} \approx 
\chi^{(-)}_{q_b}({\bf r}_{bt}) \chi^{(-)}_{q_c}({\bf r}_{ct}).
\end{eqnarray}
This leads to the prior form DWBA $T$ -- matrix
\begin{eqnarray}
T^{(-)[prior]}_{fi}(DWBA) = 
\langle \chi^{(-)}_{q_b}({\bf r}_{bt}) \chi^{(-)}_{q_c}({\bf r}_{ct})
|V_{ct} + V_{bt} -U_{at}| 
\chi^{(+)}_{q_a}({\bf r}_{at}) \phi_a({\bf r}_{bc})\rangle. 
\end{eqnarray}

It can be shown {\cite {huby}} that the 
DWBA $T$ -- matrices given by Eqs. (22) and (24) 
are equivalent to one another, i.e.,
\begin{eqnarray}
T^{(+)[post]}_{fi}(DWBA) = T^{(-)[prior]}_{fi}(DWBA).
\end{eqnarray}
Thus, for actual calculations, one may use the $T$ -- matrix
 which seems more convenient.
$T^{(-)[prior]}$ involves very complicated coordinate 
transformations as compared to the $T^{(+)[post]}$ form and hence it  
is very difficult to work with it in actual problems. Moreover $V_{bc}$ in 
Eq. (22) is of a shorter range than $V_{ct} + V_{bt} -U_{at}$ in Eq. (24),
which would make the numerical evaluation of Eq. (22) relatively easier.
The post form DWBA has been extensively used to perform breakup 
calculations {\cite {baur84,baur_dw,pam_etal,shyam_dw}}.  

However, by introducing a different approximation for $\Psi_f^{(-)}$ 
an alternate prior form $T$ -- matrix can be obtained. If we assume
 that the final state interaction $V_{bc}$ between the fragments 
$b$ and $c$ is important, then one can approximate $\Psi_f^{(-)}$ as
\begin{eqnarray}
\Psi_f^{(-)} \approx \chi^{(-)}_{{\bf Q}_f}({\bf r}_{at})
\phi^{(-)}_{a^*,{\bf q}_f}({\bf r}_{bc}).
\end{eqnarray}
In Eq. (26), the
relative motion wave function of $b$ and $c$ (which could also be a 
resonant state)
is described by $\phi^{(-)}_{a^*,{\bf q}_f}({\bf r}_{bc})$, where ${\bf q}_f$
denotes the relative momentum between the fragments. 
The center of mass (c.m.) motion of the unbound system ($a^*=b+c$) 
with respect to the target
in the final state is given by $\chi^{(-)}_{{\bf Q}_f}({\bf r}_{at})$ with
momentum ${\bf Q}_f$. They are related to momenta ${\bf q}_b$
and ${\bf q}_c$ of fragments $b$ and $c$ by
\begin{eqnarray}
{\bf Q}_f = {\bf q}_b + {\bf q}_c
\end{eqnarray}
and
\begin{eqnarray}
{\bf q}_f = {m_b \over {m_a}}{\bf q}_b - {m_c \over {m_a}}{\bf q}_c,
\end{eqnarray}
respectively.
 
This approximation [Eq. (26)] leads to an alternate prior form 
$T$ -- matrix 
\begin{eqnarray}
T^{(-)[alt,prior]}_{fi}(DWBA) = 
\langle \chi^{(-)}_{{\bf Q}_f}({\bf r}_{at})
\phi^{(-)}_{a^*,{\bf q}_f}({\bf r}_{bc})
|V_{ct} + V_{bt} -U_{at}| 
\chi^{(+)}_{q_a}({\bf r}_{at}) \phi_a({\bf r}_{bc})\rangle 
\end{eqnarray}
or
\begin{eqnarray}
T^{(-)[alt,prior]}_{fi}(DWBA) = 
\langle \chi^{(-)}_{{\bf Q}_f}({\bf r}_{at})
\phi^{(-)}_{a^*,{\bf q}_f}({\bf r}_{bc})
|V_{ct} + V_{bt}| 
\chi^{(+)}_{q_a}({\bf r}_{at}) \phi_a({\bf r}_{bc})\rangle. 
\end{eqnarray}
In Eq. (29), $U_{at}$ depends on ${\bf r}_{at}$ while
$\phi^{(-)}_{a^*,{\bf q}_f}({\bf r}_{bc})$ and $\phi_a({\bf r}_{bc})$ 
depends on ${\bf r}_{bc}$, and hence 
the explicit dependence on $U_{at}$ in Eq. (30) 
has dropped out because of the 
orthogonality of $\phi^{(-)}_{a^*,{\bf q}_f}({\bf r}_{bc})$ and 
$\phi_a({\bf r}_{bc})$. 
We note that Eq. (30) contains only two distorted
waves as against the three distorted waves in $T^{(+)[post]}$. 
The distorted waves being oscillatory even at large distances,
any reduction in their number will accelerate the convergence of the
$T$ -- matrix.
The alternate prior form 
has been used for performing breakup calculations by several authors 
 {\cite{ry_etal,shyam_prior}. It should be noted 
that $T^{(-)[alt,prior]}_{fi}(DWBA)$ is no longer
equivalent to $T^{(+)[post]}_{fi}(DWBA)$.

The $T$ -- matrix [Eq. (30)] describes a situation in which the projectile
$a$ is inelastically excited from the ground state to its continuum. If we
ignore the nuclear interactions in the distorted waves in both incident and 
final channels and also ignore the nuclear parts of interactions 
$V_{ct}$ and $V_{bt}$, then the alternate prior form $T$ -- matrix describes
the Coulomb excitation of the projectile. The semi-classical counterpart
of Eq. (30) is the Alder--Winther theory of Coulomb excitation
{\cite{alder}}.
 
The alternate prior form DWBA can be regarded as the first 
iteration of the solutions of a coupled channels problem
e.g., the coupled discretized continuum channels or CDCC equations
(see, e.g. \cite{ijt99}). However breakup studies of both 
stable isotopes {\cite{kami,baur83}}
and halo nuclei {\cite{toste01,mort,moro}} have shown that 
the alternate prior form DWBA is insufficient to
describe the data and that higher order coupling effects of the breakup 
channels are important in both the cases. 

In the next sub-section, we show that by suitably rewriting the
post form DWBA $T$ -- matrix [Eq. (22)], the features of the
spectator-participant mechanism can be explicitly seen therein
\cite{baur_chew}.

%%%%%%%%%%%%%%%%%%%%%%%%%%%%%%%%%%%%%%%%%%%%%%%%%%%%%%%%%%
\subsection{Post form DWBA $T$ -- matrix in quasi free limit}
%%%%%%%%%%%%%%%%%%%%%%%%%%%%%%%%%%%%%%%%%%%%%%%%%%%%%%%%%%
In Eq. (22), we replace the relative motion wave function of the projectile 
 by a plane wave, i.e.,
\begin{eqnarray}
\chi^{(+)}_{q_a}({\bf r}_{at}) = e^{i{\bf q}_a.{\bf r}_{at}},
\end{eqnarray}
and rewrite the scattering wave functions in the final state, in terms of 
the half
off-shell $t$-matrix elements $[t_{jt}({\bf p},{\bf q}_j), E=p^2/2m_j]$ as, 
\begin{eqnarray}
\chi^{(-)}_{q_j}({\bf r}_{jt}) = e^{i{\bf q}_j.{\bf r}_{jt}} + 
\lim_{\epsilon \rightarrow 0} {1 \over {(2\pi)^3}} \int d^3p
{{t_{jt}({\bf p},{\bf q}_j)e^{i{\bf p}.{\bf r}_{jt}}}\over {p^2-q_j^2 +i\epsilon}},
~j = b,c.
\end{eqnarray}
The $T$ -- matrix then has
four terms: (1) a term containing three plane waves,
which vanishes because of energy conservation, (2) a term ($T_c$) describing
the scattering of particle $c$ on $t$,
\begin{eqnarray}
 T_c = {1 \over {(2\pi)^3}}\lim_{\epsilon \rightarrow 0}
 \int \int \int d^3 r_{bc} d^3 r_{at} d^3 p 
{{e^{-i{\bf p}.{\bf r}_{ct}}}\over {p^2-q_c^2 -i\epsilon}} 
t^*_{ct}({\bf p},{\bf q}_c)
e^{-i{\bf q}_b.{\bf r}_{bt}} V_{bc}(r_{bc}) e^{i{\bf q}_a.{\bf r}_{at}}
\phi_a({\bf r}_{bc}), \nonumber \\
~~
\end{eqnarray}
(3) a term ($T_b$) [which is similar to that given by Eq. (33)] describing
the scattering of particle $b$ on $t$,
(4) and a term ($T_{bc}$) describing the scattering of both $b$
and $c$ on $t$ (double scattering).

Let us consider, e.g., the term $T_c$ [Eq. (33)]. 
We express ${\bf r}_{ct}$ and
${\bf r}_{bt}$ in terms of ${\bf r}_{bc}$ and ${\bf r}_{at}$ (Fig. 2) as,
\begin{figure}[ht]
\begin{center}
\mbox{\epsfig{file=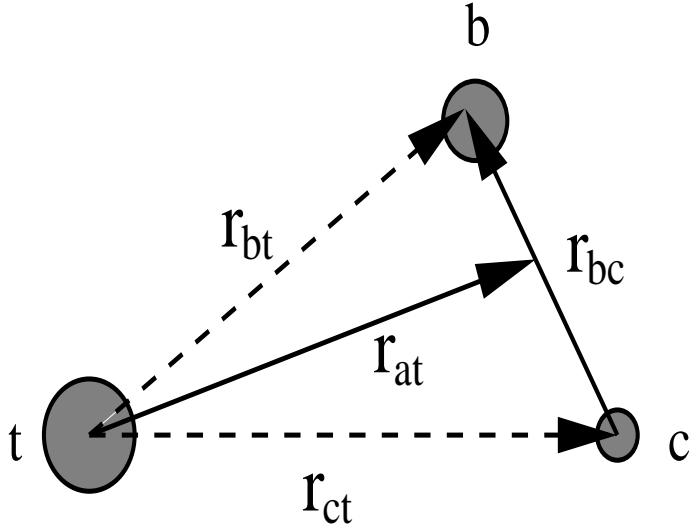,height=6.95cm,width=.75\textwidth}}
\end{center}
\caption{The coordinate system used in section 2.2. } 
\label{fig:c2_4)}
\end{figure}
\begin{eqnarray}
{\bf r}_{ct} = {\bf r}_{at} - {m_b \over {m_a}} {\bf r}_{bc}
\end{eqnarray}
and 
\begin{eqnarray}
{\bf r}_{bt} = {\bf r}_{at} + {m_c \over {m_a}} {\bf r}_{bc}.
\end{eqnarray}
Substituting Eqs. (34) and (35) in Eq. (33) we get,
\begin{eqnarray}
 T_c &=& {1 \over {(2\pi)^3}}\lim_{\epsilon \rightarrow 0}
 \int \int \int d^3 r_{bc} d^3 r_{at} d^3 p 
{{e^{i({\bf q}_a-{\bf q}_b-{\bf p}).{\bf r}_{at}}}\over {p^2-q_c^2 -i\epsilon}} 
e^{i({m_b \over {m_a}} {\bf p}-{m_c \over {m_a}}{\bf q}_b).{\bf r}_{bc}} 
t^*_{ct}({\bf p},{\bf q}_c) \nonumber \\
& \times & V_{bc}(r_{bc}) \phi_a({\bf r}_{bc})
\end{eqnarray} 
Integrations over coordinates ${\bf r_{at}}$ and ${\bf p}$ can be performed
easily to give,
\begin{eqnarray}
T_c&=& 
{{t^*_{ct}({\bf q}_a-{\bf q}_b,{\bf q}_c)}\over {({\bf q}_a-{\bf q}_b)^2-q_c^2}}
F({\bf Q}), 
\end{eqnarray}
where
\begin{eqnarray}
F({\bf Q}) = \int d^3 r_{bc} 
{{e^{i{\bf Q}.{\bf r}_{bc}}}}
V_{bc}(r_{bc}) \phi_a({\bf r}_{bc}),
\end{eqnarray}
with
\begin{eqnarray}
{\bf Q} = {m_b \over {m_a}} {\bf q}_a-{\bf q}_b.
\end{eqnarray}
Now the Schr\"{o}dinger equation for the $b-c$ system can be used to 
rewrite Eq.~(38) as 
\begin{eqnarray}
F({\bf Q}) 
 &=& -(2\pi)^{3/2} {{\hbar^2}\over {2\mu_{bc}}} (\alpha^2 + Q^2)
 {\tilde \phi}_a({\bf Q}),
\end{eqnarray}
where
\begin{eqnarray}
{\tilde \phi}_a({\bf Q})=(2\pi)^{-3/2}\int d^3 r_{bc} e^{i{\bf Q}.{\bf r}_{bc}}
\phi_a({\bf r}_{bc})
\end{eqnarray}
is the momentum space wave function of the projectile in its ground state. 
In Eq.~(40) $\alpha$ is defined as 
\begin{eqnarray}
\epsilon_{bc} = {{\hbar^2 \alpha^2}\over {2\mu_{bc}}},
\end{eqnarray}
where $\epsilon$ is the binding energy of the $b-c$ system
and $\mu_{bc}$ is the corresponding reduced mass. 
Thus $T_c$ is written as
\begin{eqnarray}
T_c = -(2\pi)^{3/2} {{\hbar^2}\over {2\mu_{bc}}}
{{t^*_{ct}({\bf q}_a-{\bf q}_b,{\bf q}_c)}\over {({\bf q}_a-{\bf q}_b)^2-q_c^2}}
(\alpha^2 + Q^2) {\tilde \phi}_a({\bf Q}).
\end{eqnarray}
Now, for an infinitely heavy target, the energy conservation gives
\begin{eqnarray}
-{{(\alpha^2 + Q^2)}\over {({\bf q}_a-{\bf q}_b)^2-q_c^2}} = {m_b \over {m_a}}.
\end{eqnarray}
Substituting Eq. (44) in Eq. (43) we obtain,
\begin{eqnarray}
 T_c = {\hbar^2 \over{2m_c}}(2\pi)^{3/2}t^*_{ct}({\bf q}_a-{\bf q}_b,{\bf q}_c)
 {\tilde \phi}_a({\bf Q}).
\end{eqnarray}
A similar expression holds for $T_b$. 

If the off-shell dependence of 
$t^*_{ct}({\bf q}_a-{\bf q}_b,{\bf q}_c)$ in Eq. (45), is neglected 
{\cite{baur_chew}}, then 
the breakup cross section is determined by the modulus square of the 
wave function, ${\tilde \phi}_a({\bf Q})$. The maximum cross section would
occur when
\begin{eqnarray}
{\bf Q} = {m_b\over m_a}{\bf q}_a -{\bf q}_b =0,
\end{eqnarray}
or
\begin{eqnarray}
{\bf v}_a = {\bf v}_b,
\end{eqnarray}
i.e., when particle $b$ moves with the beam velocity and
is only a `spectator' in the breakup process.
Thus, the term $T_c$ (or $T_b$) has the simple physical interpretation: particle 
$c$ (or $b$) interacts with $t$ via $V_{ct}$ (or $V_{bt}$) and is 
knocked out of the projectile, while particle $b$ (or $c$) is only a
spectator. This shows that the main feature of the Serber model -- the
dependence of the cross section on the internal momentum distribution of the
fragment within the projectile -- is embedded within the post form 
DWBA $T$ -- matrix in a quasi free limit. 

It must, however, be cautioned
that while some gross physical insights are obtained from the quasi free limit
of the post form DWBA $T$ -- matrix, actual calculations must be made with
full distorted waves {\cite{shyam79}} in order to explain the breakup data.

The breakup reactions of halo nuclei have been investigated by several
authors using a variety of approaches (see, e.g., \cite{esbe95,mele99,bon00,
yab92}, and \cite{baur03}, for an exhaustive bibliography on the subject).
However, only a few \cite{shyam_prior,typ01,raj02,raj03,mar02} of them treat
both Coulomb and nuclear breakup terms consistently on an equal footing.  
In Ref.~\cite{shyam_prior} the prior form $T$ -- matrix given by Eq. (30)
has been used to study the breakup reactions of $^8$B. It should be
recalled that in this $T$ -- matrix the fragment-target interaction is treated 
in first order which was subsequently shown to be inadequate. In
Refs.~\cite{typ01,mar02} the time evolution of the projectile in the
coordinate space is described by solving the time dependent Schr\"odinger
equation, treating the projectile target interaction as a time dependent
external perturbation. These calculations use the semiclassical concept
of the motion of the projectile along a trejectory.

The post form DWBA formulation of breakup reactions which uses the $T$
matrix given by Eq.~(20) includes consistently both Coulomb and nuclear
interactions between the projectile fragments and the target nucleus to
all orders, but treats the fragment-fragment interaction in first order.
As can be noted easily, the full wave function describing the ground
state structure of the projectile, enters as an input in this theory
which makes it possible to investigate the structure of the
projectile from the study of the breakup reactions. It can treat the
Coulomb and nuclear breakups as well as their interference terms
consistently on an equal footing. Since this theory uses the post
form scattering amplitude, the breakup contributions from the entire
valence nucleon-core fragment continuum corresponding to all the
multipoles and the relative angular momenta are included in it. This
can account for the postacceleration effects in a unique way. In the
subsequent sections we describe this theory in some details.

%%%%%%%%%%%%%%%%%%%% E2 %%%%%%%%%%%%%%%%%%%%%%%%%%%%%%%%%%%%%%%%%
%%%%%%%%%%%%%%%%%%%% B3 %%%%%%%%%%%%%%%%%%%%%%%%%%%%%%%%%%%%%%%%%
\section{Breakup amplitudes in the post form DWBA}

\subsection{Pure Coulomb breakup}  
 
Coulomb dissociation of halo nuclei has been investigated by several
authors using a number of different theoretical 
approaches. A semiclassical coupled channels formalism has been used 
by authors of Ref. \cite{cant93}, while in Refs. \cite{kid94,ber93} 
the time dependent Schr\"odinger equation method
was employed. The results within these approaches depend on the range
of the impact parameter associated with the straight line trajectories
used to describe the motion of the projectile in the field of target
nuclei. However, in these studies the emphasis was on investigating the
dynamics of the Coulomb dissociation, and not the structure
of the projectile ground state which was assumed to have some very
simple zero range (ZR) form. Similar assumption for the projectile
structure was also made in other semiclassical~\cite{han96_3,ann94} and
prior form distorted wave Born approximation (DWBA) calculations~\cite{bert91}.
 
In this section, we present a theoretical model to describe the 
pure Coulomb breakup of one-neutron halo nuclei within the framework
of the post form DWBA where finite range effects are included via a
local momentum approximation (LMA)~\cite{braun74a,braun74b,shyam85}.
This theory of breakup reactions
incorporates the details of the ground state structure of the projectile 
in the breakup amplitude~\cite{baur84,raj00}.

We consider the reaction $ a + t \rightarrow b + c + t $, where the 
projectile $a$ breaks up into fragments $b$ (charged) 
and $c$ (uncharged) in the Coulomb
field of a target $t$. The coordinate system chosen is shown in Fig. 3.
\begin{figure}[ht]
\begin{center}
\mbox{\epsfig{file=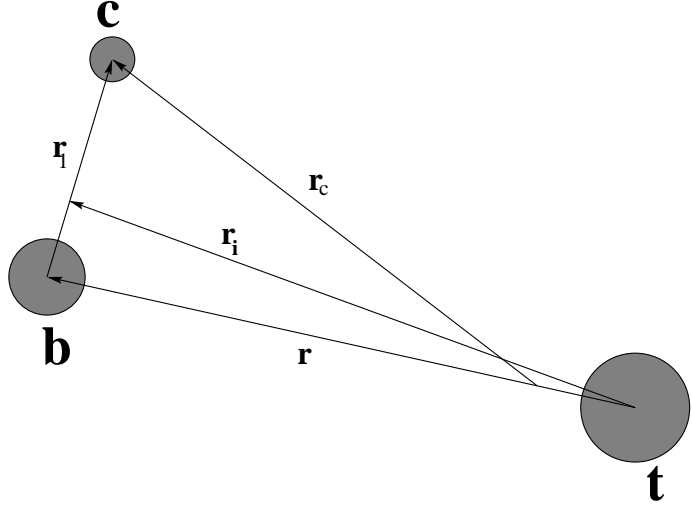,width=.75\textwidth}}
\end{center}
\caption{Coordinate system used in the $T$-matrix (Eq. 50). 
$b$, $c$ and $t$ represent the
charged core, valence neutron and target, respectively. } 
\label{fig:figa)}
\end{figure}

The position vectors satisfy the following relations
\begin{eqnarray}
{\bf r} &=& \ri - \alpha\ro,~~ \alpha = {m_c\over {m_c+m_b}}   \\
\rc &=& \gamma\ro +\delta\ri, ~~ \delta = {m_t\over {m_b+m_t}}, 
~~  \gamma = (1 - \alpha\delta)    
\end{eqnarray}
The starting point of our discussion is the post form DWBA $T$ - matrix
[Eq. (20)] 
\begin{eqnarray}
 T  = \int\int\int && d\xi d\ro d\ri \cm_b(\bq,{\bf r})\Phi^*_b(\xi_b)  
 \cm_c{\cq,\rc}\Phi^*_c(\xi_c)\vv  \nonumber  \\
 &\times& {}\Phi_a(\xi_a,\ro)\cp_a(\ak,\ri),   
\end{eqnarray}
where $\chi's$ are the distorted waves for relative motions of 
$b$ and $c$ with respect to $t$ and the center of mass (c.m.) of the $b-t$
system, respectively, and $\Phi's$ are the internal state wave functions
of concerned particles with internal coordinates $\xi$. 
$\bq$, $\cq$ are Jacobi wave vectors of $b$
and $c$, respectively in the final channel of the reaction. $\vv$ is the 
interaction between $b$ and $c$. The charged fragment $b$ and the 
projectile $a$ interacts with the target by a point Coulomb interaction
and hence the corresponding distorted waves are the 
Coulomb distorted waves with appropriate boundary  
conditions. For pure Coulomb breakup, the interaction between
the target and uncharged fragment $c$ is zero and hence
for this case we can write $\cm_c{\cq,\rc} = e^{-i(\cq.\rc)}$. 
$\Phi_a(\xi_a,\ro)$ represents the bound state
wave function of the projectile having its radial and angular parts
as $u_{\ell}(r_1)$ and ${Y^{\ell}_{m}({\hat{\bf r}}_1)}$,
respectively, which are associated with the relative motion of $b$ and $c$.  
Integrals over the internal coordinates $\xi$ give
\begin{eqnarray}
 &&\int d\xi\Phi^*_b(\xi_b)\Phi^*_c(\xi_c)\Phi_a(\xi_a,\ro)  \nonumber \\
 &&= \sum_{\ell mj\mu} \langle \ell mj_c\mu_c|j\mu\rangle 
 \langle j_b\mu_bj\mu|j_a\mu_a\rangle i^\ell u_\ell(r_1)
 {Y^{\ell}_{m}({\hat{\bf r}}_1)}.
\end{eqnarray}
In Eq. (51), $\ell$ is the orbital angular momentum
for the relative motion between $b$ and $c$, 
$j_a$ is the spin of $a$. Using this equation
the $T$-matrix can be written as  
\begin{eqnarray} 
 T & = & \sum_{\ell mj\mu} \langle \ell mj_c\mu_c|j\mu\rangle 
 \langle j_b\mu_bj\mu|j_a\mu_a\rangle i^\ell
 \hat{\ell}\beta_{\ell m}(\bq,\cq;\ak),
\end{eqnarray}
where
\begin{eqnarray}
\hat{\ell}\beta_{\ell m}(\bq,\cq;\ak)  =  
\int\int && d\ro d\ri\cm_b(\bq,{\bf r})e^{-i\cq.\rc}
 \vv \nonumber \\
 && \times {} \phi^{\ell m}_a(\ro) \cp_a(\ak,\ri),    
\end{eqnarray}
with $\beta_{\ell m}$ being the reduced $T$ -- matrix and
 ${\hat \ell} = \sqrt{2\ell + 1}$. We have 
written $\phi^{\ell m}_a(\ro) = u_\ell(r_1){Y^{\ell}_{m}({\hat{\bf r}}_1)}$. 

It may be noted that the reduced amplitude $\beta_{\ell m}$ involves
a six dimensional integral which makes its evaluation quite complicated.
The problem gets further acute due to the fact that the integrand involves
three scattering waves which have oscillatory behaviour asymptotically.
Therefore, several approximate methods have been used in the literature
to avoid the evaluation of six dimensional integrals.
 In the zero range approximation
(ZRA) \cite{satchler} one assumes 
\begin{eqnarray}
 \vv\phi^{\ell m}_a(\ro) & = & D_{0}\delta(\ro),
\end{eqnarray}
 where $D_0$ is the usual zero range constant. This approximation 
reduces the six dimensional integral in Eq. (53) to a three-dimensional 
one. The corresponding amplitude is written as 
\begin{eqnarray}
\beta^{ZR}_{00} & = & D_{0}
\langle \chi^{(-)}_b(\bq,\ri) e^{i\delta\cq.\ri}|\cp_a(\ak,\ri)\rangle.
\end{eqnarray}
In Eq. (55), the details of the projectile structure
enter in the reaction amplitude only as a multiplicative constant $D_0$. 
However, ZRA necessarily restricts the relative
motion between $b$ and $c$ in the projectile $a$ to $s$ -- state only.
Even for such cases, this approximation may not be satisfied for
heavier projectiles and at higher beam energies \cite{shyam85}.
 
Baur and Trautmann (BT) \cite{BT} have proposed an alternative approximation
in which the projectile c.m. coordinate in the corresponding 
distorted wave in Eq. (53) is replaced by that of the core-target system, 
i.e. $\ri \approx {\bf r}$. With this approximation the amplitude
$\beta_{\ell m}$ splits into two factors each involving a three dimensional
integral 
\begin{eqnarray}
{\hat \ell}\beta^{BT}_{\ell m} & = & {\langle e^{i\cq.\ro}|V_{bc}|
\phi^{\ell m}_a(\ro)\rangle}
\langle \chi^{(-)}_b(\bq,{\bf r}) e^{i\delta\cq.{\bf r}}|\cp_a(\ak,{\bf r})
\rangle.
\end{eqnarray}
The first term depends on the structure of
the projectile through its ground state wave function
 $\phi^{\ell m}_a({\bf r_1})$. 
The second term involves the dynamics of the reaction, which can be 
expressed analytically in terms of the bremsstrahlung integral {\cite{bem}}.
This amplitude (which will be referred to as the BT amplitude),
used originally to study the deuteron breakup at 
sub-Coulomb energies \cite{BT}, was applied to calculations of
the Coulomb breakup of halo nuclei in Ref. \cite{shyam92}. This approximation,
which allows the  
application of the theory to non-$s$ -- wave projectiles,   
may seem to be justified if the c.m of the $b-c$ system
is shifted towards $b$ (which is indeed the case if $m_b \gg m_c$). 
However, the neglected piece of ${\bf r}_i$ ($\alpha \ro$) occurs in   
association with the wave vector $\ak$, whose magnitude
could be quite large for the reactions at higher beam energies. 
Therefore, contributions coming to the amplitude from the neglected part of
${\bf r}_i$ may still be substantial. 
     
An approximate way of taking into account the finite range effects in
the post form DWBA theory is provided by the local momentum approximation
\cite{shyam85,braun74a}. The attractive feature of this approximation
is that it leads to the factorization of the amplitude $\beta_{\ell m}$
similar to that obtained in the BT case.  We use this
approximation to write the Coulomb distorted wave of particle $b$ in the 
final channel as 
\begin{eqnarray}
\chi^{(-)}_b(\bq,{\bf r}) & = & e^{-i\alpha{\bf K}.\ro}
                           \chi^{(-)}_b(\bq,\ri). 
\end{eqnarray}
Eq. (57) represents an exact Taylor series expansion about ${\bf r}_i$ if 
${ {\bf K}}( = -i\nabla_{{\bf r}_i})$ is treated exactly. However,
this is not done in the LMA scheme. Instead, the
magnitude of the local momentum is taken to be 
\begin{eqnarray} 
{ {K}}(R) = {\sqrt {{2m\over \hbar^2}[E - V(R)]}},
\end{eqnarray}
where $m$ is the reduced mass of the $b-t$ system,
$E$ is the energy of particle $b$ relative to the target in the
c.m. system and $V(R)$ is the  Coulomb potential
between $b$ and the target at a distance $R$. Thus, 
the local momentum ${{\bf K}}$ is evaluated at some distance
$R$, and its magnitude is held fixed for all the values of ${\bf r}$.
As shown in appendix of Ref. \cite{raj00}, the magnitude of ${\bf K}$ remains constant
for ${\bf r} > 10$ fm. Due to the peripheral nature of breakup reactions, this
region contributes maximum to the cross section. Therefore, we have 
taken a constant magnitude for ${\bf K}$ evaluated at $R=10$ fm for 
all the values of $r$. As is discussed in \cite{raj00,raj03} the results
of our calculations are almost independent of the choice of the direction
of the local momentum. Therefore, we have taken the directions of
${{\bf K}}$ and ${\bf k_b}$ to be the same in all the calculations.
 Detailed discussion on the validity of the
local momentum approximation is presented in Refs.~\cite{raj00,raj03}. 
It may be noted that in the calculations presented in  Ref. \cite{pb1},
the LMA was applied to the Coulomb distorted wave of the projectile 
channel which could imply a deviation from the distorted wave approximation.

Substituting Eq. (57) into Eq. (53) we get the following factorized form of
the reduced amplitude  
\begin{eqnarray}
{\hat \ell}\beta^{FRDWBA} _{\ell m}& = &
\langle e^{i(\gamma\cq - \alpha {\bf K}).\ro}|V_{bc}|
\phi^{\ell m}_a(\ro)\rangle \nonumber \\
        & \times &\langle \chi^{(-)}_b(\bq,\ri) e^{i\delta\cq.\ri}|
\cp_a(\ak,\ri)\rangle.
\end{eqnarray}
Eq. (59) (which will be referred to as the FRDWBA amplitude in the following) 
looks like Eq. (56) of the BT theory but with the
very important difference that the  
form factor is now evaluated at the momentum transfer  
($\gamma {\bf k}_c - \alpha{ {\bf K}}$), and not at the valence particle 
momentum ${\bf k_c}$. The two momenta could be quite different for 
cases of interest in this work.  
The second term, involving the dynamics of the reaction, is the same
in both the cases. Therefore, the breakup amplitude obtained in 
BT approximation differs from that of FRDWBA by a factor 
\begin{eqnarray}
F_r  = {{\beta^{BT}_{\ell m}}\over {\beta^{FRDWBA}_{\ell m}}} = 
{{\langle e^{i\cq.\ro}|V_{bc}|\phi^{\ell m}_a(\ro)\rangle}\over
{\langle e^{i(\beta\cq - \alpha {\bf K}).\ro}|V_{bc}|\phi^{\ell m}_a(\ro)\rangle}}
\end{eqnarray}
\begin{figure}[ht]
\begin{center}
\mbox{\epsfig{file=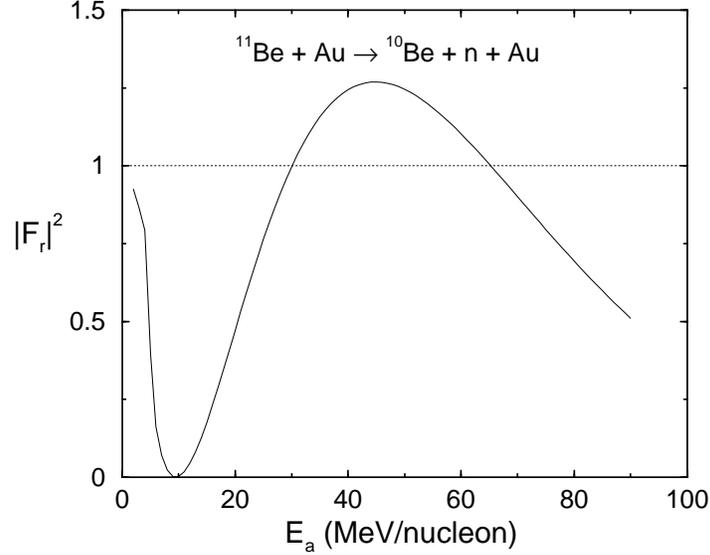,width=.75\textwidth}}
\end{center}
\caption{Modulus square of the ratio $F_r$ defined by Eq. (60) as a 
function of the beam energy for the breakup of $^{11}$Be on a gold
target at the beam energy of 44 MeV/nucleon corresponding to forward
emission angles of the breakup  fragments as discussed in the text.}
\label{fig:figb}
\end{figure}

In Fig. 4, we have shown the
beam energy dependence of $|F_r|^2$ for the 
breakup of $^{11}$Be on a gold target at the beam energy of
44 MeV/nucleon for a set of forward angles of the outgoing
fragments ($\theta _b = 1^{\circ}$, $\theta _c = 1^{\circ}$ and 
$\phi _c = 1^{\circ}$). In this calculation, the ground state
wave function of $^{11}$Be has been obtained by  considering a
configurations in which  a $s$ -- wave valence neutron is
coupled to $0^+$  $^{10}$Be core [$^{10}$Be$(0^+) \otimes 1s_{1/2}\nu$]
with the one-neutron separation energy ($S_{n-^{10}Be}$) of 504 keV
and a spectroscopic factor (SF) of 0.74 \cite{jon,auton70,zwieg79}.
The single particle wave function has been obtained by assuming
the valence neutron-$^{10}$Be interaction to be of Woods-Saxon type
whose depth is adjusted to reproduce the corresponding binding energy
with fixed values of the radius and diffuseness parameters
(taken to be  1.15 fm and 0.5 fm, respectively)(see \cite{raj00} for
more details). In this figure, we see that $|F_r|^2$ is close 
to unity only at the sub-Coulomb beam energies (of course at
higher beam energies it crosses twice the line representing the value 1).
Therefore, the BT and FRDWBA calculations are expected to produce
similar results at very low incident energies. Depending upon
the beam energy, the BT results can be larger or smaller than 
those of the FRDWBA.

Recently, a theory of the Coulomb breakup has been developed within an 
adiabatic (AD) model \cite{toste98,jal97}, where one assumes that the 
excitation of the projectile is such that the relative energy
($\epsilon_{bc}$) of the $b-c$ system is much smaller than the total
incident energy, which allows $\epsilon_{bc}$ to be replaced by the
constant separation energy of the fragments in the projectile ground
state. It was shown in \cite{jal97} that under these conditions the 
wave function $\we$ has an exact solution as given below
\begin{eqnarray}
\Psi_a^{(+),AD}(\xi_a,{\bf r}_1,{\bf r}_i) & = &
 \Phi_a(\xi_a,\ro)e^{i\alpha\ak.\ro}\cp_a(\ak,{\bf r}).
\end{eqnarray}

Substituting $\Psi_a^{(+),AD}$ for $\Psi_a^{(+)}$ in Eq. (50)
leads to the reduced amplitude:
\begin{eqnarray}
{\hat \ell}\beta^{AD}_{\ell m} & = &
\langle e^{i(\cq - \alpha\ak).\ro}|V_{bc}|\phi^{\ell m}_a(\ro)\rangle
\langle \chi^{(-)}_b(\bq,{\bf r}) e^{i\delta\cq.{\bf r}}|\cp_a(\ak,{\bf r})
\rangle.
\end{eqnarray}
This amplitude differs from those of BT as well as  FRDWBA only in
the form factor part (first term), which is evaluated here at the momentum 
transfer $(\cq - \alpha \ak)$. Eq. (62) can also be obtained in the
DWBA model by making a local momentum approximation to the Coulomb
distorted wave in the initial channel of the reaction, and by evaluating
the local momentum at $R = \infty$ with its direction being the
same as that of the projectile \cite{pb1}. The adiabatic model does
not make the weak coupling approximation of the DWBA. However,
it necessarily requires one of the fragments (in this case $c$) to be
chargeless. In contrast, the FRDWBA can be applied to cases where both
the fragments $b$ and $c$ are charged (see, e.g., Ref. \cite{shyam85}).
While the effect of nuclear breakup in the adiabatic model description of the 
elastic scattering of the loosely bound projectile has been calculated in 
Refs. \cite{jal97,jc97,ron98}, the nuclear part of the amplitude for  breakup
reactions is yet to be calculated within this model. However, 
the calculations of the nuclear breakup cross section has been done within the 
FRDWBA theory \cite{raj03}. 

The triple differential cross section of the reaction is 
given by \begin{eqnarray}
{{d^3\sigma}\over{dE_bd\Omega_bd\Omega_c}} & = & 
 {2\pi\over{\hbar v_a}}\rho(E_b,\Omega_b,\Omega_c)\sum_{\ell m}
|\beta_{\ell m}|^2,
\end{eqnarray}
where $\rho(E_b,\Omega_b,\Omega_c)$ is the appropriate 
\cite{fuchs} three-body phase space factor, given by 
\begin{eqnarray}
\rho(E_b,\Omega_b,\Omega_c) & = &{h^{-6}m_bm_cm_tp_bp_c \over 
m_t+m_c-m_c{{\bf k_c.(k_a-k_b)} \over k_c^2}}, 
\end{eqnarray}
with ${\bf k}_a,{\bf k}_b$ and ${\bf k}_c$ being evaluated in
the appropriate frame of reference. $v_a$
is the relative velocity of the projectile in the initial channel.
In Eq. (64), the linear momenta, $p$, are related to wave numbers $k$
by $p \,\, = \,\, \hbar k$.

Substituting the following expressions for the Coulomb distorted waves
\begin{eqnarray}
\cm_b(\bq,{\ri}) & = &
 e^{-\pi\eta_b/2}\Gamma(1 + i\eta_b) e^{-i\bq.\ri} 
 {_1F_1(-i\eta_b, 1, i(k_b r_i + \bq.\ri))},   \\
                                          \nonumber \\
\cp_a(\ak,\ri) & = &
 e^{-\pi\eta_a/2}\Gamma(1 + i\eta_a) e^{i\ak.\ri} 
 {_1F_1(-i\eta_a, 1, i(k_a r_i - \ak.\ri))}
\end{eqnarray}
in Eq. (59), one gets for the triple differential cross section
\begin{eqnarray}
{{d^3\sigma}\over{dE_bd\Omega_bd\Omega_c}} = {2\pi\over{{\hbar}v_a}}
\rho(E_b,\Omega_b,\Omega_c) 
{4\pi^2\eta_a\eta_b\over (e^{2\pi\eta_b}-1)(e^{2\pi\eta_a}-1)}|I|^2  
4\pi\sum_{\ell} |Z_{\ell}|^2.
\end{eqnarray}
In Eqs. (65 -- 67), $\eta$'s are the Coulomb parameters for the concerned
particles. In Eq. (67), $I$ is the bremsstrahlung integral \cite{bem}
which can be evaluated in a closed form:  
\begin{eqnarray}
I &=& -i{\Big[}B(0){\Big(}{{dD}\over{dx}}{\Big)}_{x=0}(-\eta_a\eta_b)\fa  \nonumber \\
& + & {\Big(}{{dB}\over{dx}}{\Big)}_{x=0} {\fB} {\Big]},
\end{eqnarray}
where
\begin{eqnarray}
B(x) = {4\pi\over{k^{2(i\eta_a+i\eta_b+1)}}}
{\Big[}(k^2 - 2{\bf k}.\ak -2xk_a)^{i\eta_a}
(k^2 - 2{\bf k}.\bq -2xk_b)^{i\eta_b}{\Big]},
\end{eqnarray}
\begin{eqnarray}
D(x) = {2k^2(k_ak_b+\ak.\bq)-4({\bf k}.\ak+xk_a)({\bf k}.\bq+xk_b)\over
{(k^2 - 2{\bf k}.\ak -2xk_a)(k^2 - 2{\bf k}.\bq -2xk_b)}}
\end{eqnarray}
with ${\bf k} = \ak - \bq -\delta\cq$.
$Z_{\ell}$ contains the projectile structure information and is given by
\begin{eqnarray}
Z_{\ell} = \int dr_1 r^2_1 j_{\ell} (k_1 r_1)\vv u_{\ell} (r_1),
\end{eqnarray}
with $k_1 = |\gamma\cq - \alpha {\bf K}|$.

Let us now discuss some numerical applications of the theory of
the pure Coulomb breakup reactions as presented above.
We investigate the breakup of neutron rich nuclei $^{11}$Be and
$^{15,17,19}$C at beam energies below 100 MeV/nucleon.
Apart from the distance at which the local momentum is calculated (which is
taken to be 10 fm ) and its direction
(described earlier), the only other input
to our calculations is the radial part of the projectile ground state
wave function. As discussed above, we have assumed a Woods-Saxon potential
to describe the valence neutron-core relative motion whose depth is 
searched, for a given configuration, to reproduce the corresponding
binding energy. In the calculations presented here we have mostly 
considered a $s$-wave configuration (as described above) for the 
$^{11}$Be ground state. However, in some cases we have also considered a
$d$-wave configuration for this nucleus in which a $d$-wave valence neutron   
is coupled to $2^+$  $^{10}$Be core [$^{10}$Be$(2^+) \otimes 0d_{5/2}\nu$]
with the one-neutron separation energy ($S_{n-^{10}Be}$) of 3.872 MeV
\cite{raj00}.  
The configurations for the C isotopes used in our calculations are 
described at the appropriate places below.

In Fig. 5, we present a comparison of our calculation with the data 
(taken from \cite{ann94}) for the neutron energy distribution of the 
double differential cross section ($d^2\sigma/dE_n d\Omega_n$)
at the neutron angles of $1^\circ$ and $3.4^\circ$, in the
breakup of $^{11}$Be on a gold target at the beam energy of
44 MeV/nucleon. Calculations performed within both FRDWBA and AD model
of pure Coulomb breakup are shown in this figure.
\begin{figure}[ht]
\begin{center}
\mbox{\epsfig{file=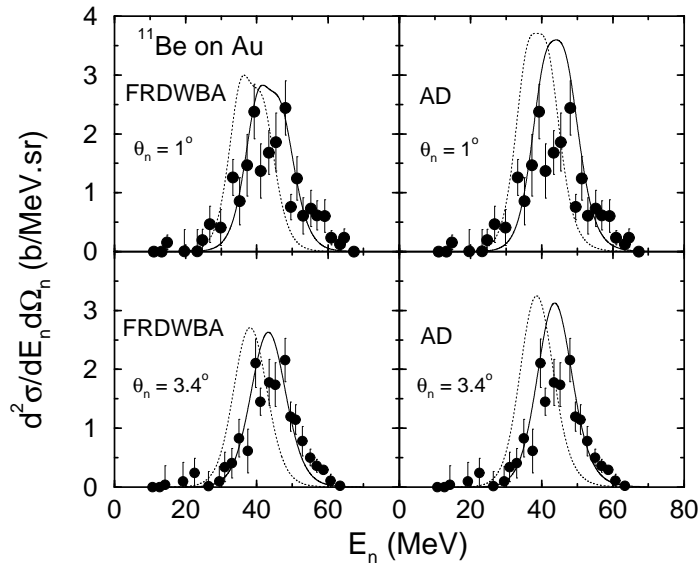,width=.75\textwidth}}
\end{center}
\caption{Neutron energy distributions for the breakup of $^{11}$Be on Au at
beam energies of 37 MeV/nucleon (dotted lines) and 44 MeV/nucleon (solid 
lines), calculated using configuration (a) with single particle 
wave functions within the FRDWBA and the AD models. The top half of the
figure is for $\theta_n = 1^{\circ}$, while the bottom half is for 
$\theta_n = 3.4^{\circ}$. The experimental data are taken from 
\protect\cite{ann94}.} 
%\label{fig:figc}
\end{figure}
The $s$-wave configuration has been used for the $^{11}$Be ground state
in both the cases. The beam energy in this experiment
\cite{ann94} varies between 36.9 -- 44.1 MeV/nucleon. To take into account
this spread, we have performed calculations at both
its upper (44 MeV/nucleon) (solid line) and lower ends
(37 MeV/nucleon) (dotted line). Even though these data have large statistical
errors, the calculations performed at 44 MeV/nucleon are in better agreement
with the experimental values. It should also be noted that the AD model 
calculations over-predict the experimental cross sections in the peak region.

The measured neutron angular distribution in the exclusive
$^{11}$Be + $A$ $\rightarrow$ $^{10}$Be + n +$A$ reaction on heavy targets below the
grazing angle is very narrow and is shown to be \cite{pb1,pb93}
dominated by the Coulomb breakup process. This reflects the narrow
width of the transverse momentum distribution of the valence neutron
in the ground state of $^{11}$Be, which is consistent with the presence
of a neutron halo structure in $^{11}$Be. In Fig. 6, we compare the
calculated and measured exclusive neutron angular distribution
$d\sigma/d\Omega_n$ as a function of the neutron angle $\theta_n$ for
the above reaction on Au, Ti and Be targets at the beam energy of
41 MeV/nucleon. The $^{11}$Be ground state wave function is the same
as described above. We note that for Au and Ti targets pure coulomb breakup
calculations are able to describe the data at forward angles (for
neutron angles below 25$^\circ$ and 15$^\circ$, respectively in the two
case). On the other hand, for the $^9$Be target pure breakup calculations
are much below the data everywhere. This gives a clear indication of the
importance of nuclear breakup effects at the backward angles for
the medium mass and heavy targets and everywhere for the light target.
\begin{figure}[ht]
\begin{center}
\mbox{\epsfig{file=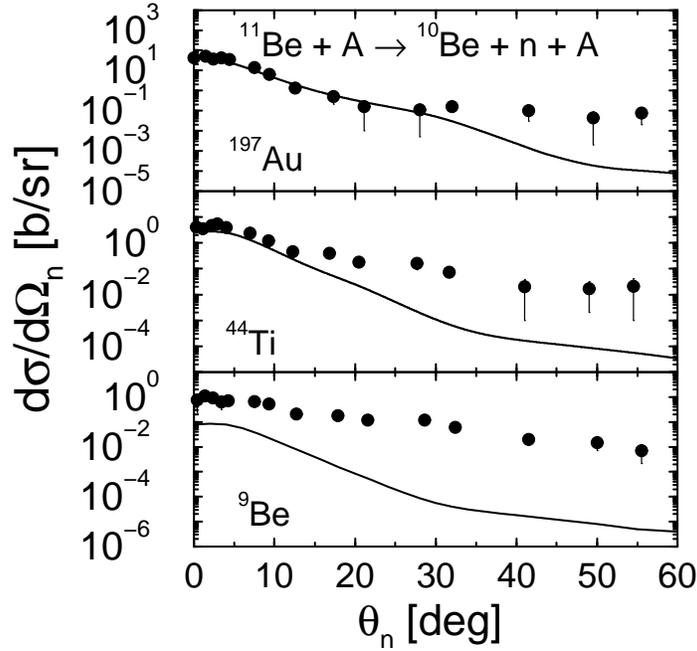,width=.75\textwidth}}
\end{center}
\caption{The calculated neutron angular distributions for 
the breakup of $^{11}$Be on a Au, Ti and Be targets at 41 MeV/nucleon
beam energy. The data are taken from \protect\cite{ann94}.}  
\label{fig:figd}
\end{figure}

Pure Coulomb contribution to the relative energy spectrum in
the breakup of $^{11}$Be on a Pb target at 72 MeV/nucleon is
shown in Fig. 7. The top half shows the results obtained with 
FRDWBA (solid line), AD model (dotted line) and the BT approximation
(dashed line) using the $s$-wave configuration for the $^{11}$Be
ground state. In the bottom half, we show additionally the results 
obtained with the $^{11}$Be wave function calculated within a dynamical
core polarization (DCP) model~\cite{len98} (dotted line) and that
obtained with the $d$-wave configuration as described above.
We see that, while both the FRDWBA and the AD model calculations reproduce
the peak value of the data \cite{nak94} well, the FRDWBA calculations done
with the DCP wave function overestimate it.  On the other hand, none of the
calculations is able to explain the data at higher relative energies. 
This can be attributed to the fact that nuclear breakup effects, which 
can contribute substantially \cite{dasso99} at higher relative energies
(for $E_{rel}$ $>$ 0.6 MeV), are not included in these calculations.
Of course, authors of Ref. \cite{nak94} claim that their data have been
corrected for these  contributions. However, the procedure adopted
by them for this purpose is inadequate. They obtained the nuclear
breakup contribution on the Pb target, by scaling the cross sections
measured on a carbon target. This scaling procedure is unlikely to be
accurate for reactions induced by halo nuclei due to the presence of
a long tail in their ground state.
\begin{figure}[ht]
\begin{center}
\mbox{\epsfig{file=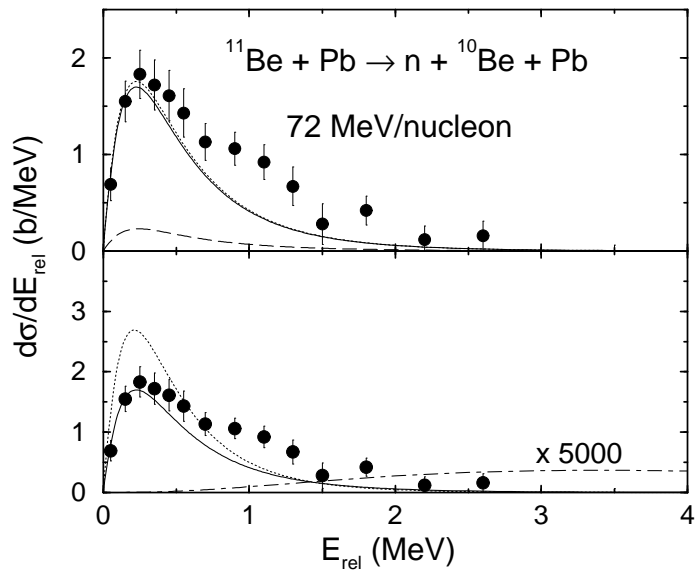,width=.75\textwidth}}
\end{center}
\caption{Relative energy spectra for the Coulomb breakup of $^{11}$Be on a
Pb target at 72 MeV/nucleon beam energy. The top half of the figure shows
the spectra obtained with a single particle wave function,
using the FRDWBA (solid line), the AD model (dotted line) and the 
BT approximation (dashed line). The bottom half shows the results of
FRDWBA calculations using single particle (solid line) and DCP
(dotted line) wave functions. The dot-dashed line shows the $d$ -- state
FRDWBA calculation using a single particle wave function,
after being multiplied by a factor of 5000. 
The data are taken from \protect\cite{nak94}.} 
\label{fig:fige}
\end{figure}

In a full quantum mechanical
theory, both Coulomb and nuclear breakup contributions should be calculated 
on the same footing and corresponding amplitudes should be added coherently.
This is discussed in the next section. 
Calculations done using the BT theory (dashed line in the upper part
of Fig. 7) underestimates the data considerably. This difference between
the FRDWBA and the BT results can again be traced to the behaviour
of $|F_r|^2$ in Fig. 4, which is smaller than unity at the beam energy
of 72 MeV/nucleon of this reaction. The FRDWBA result using the $d$ -- state 
configuration with single particle wave function (dot-dashed line
in the lower half of Fig. 7, shown after multiplying the actual numbers
by 5000) also grossly underestimates the data. 

The neutron halo structure is reflected in the narrow width of the parallel
momentum distribution (PMD) of the charged breakup fragments emitted 
in breakup reactions induced by the halo nuclei.
This is because the PMD is least affected
by the reaction mechanism \cite{orr95,mex,baz,bm92_3,ps95} and therefore, 
a narrow PMD can be related to a long tail in the  matter distribution
in the coordinate space via Heisenberg's uncertainty principle. 
\begin{figure}[ht]
\begin{center}
\mbox{\epsfig{file=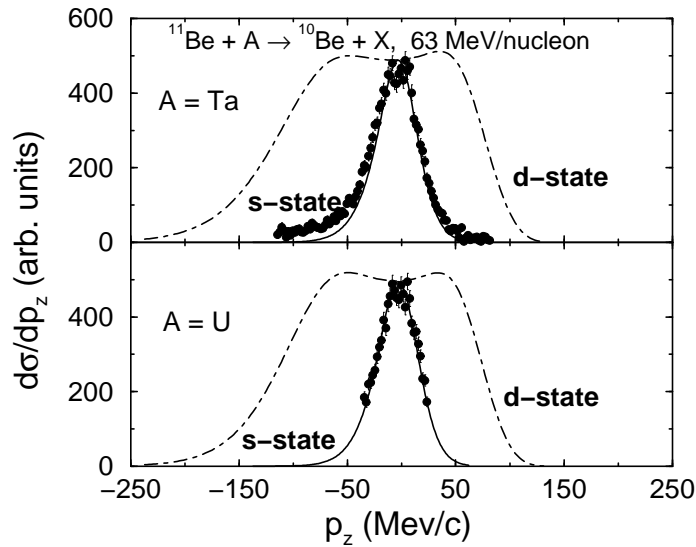,width=.75\textwidth}}
\end{center}
\vskip -0.3in
\caption{
Parallel momentum distributions of $^{10}$Be in the breakup of 
$^{11}$Be on Ta (top half) and U (bottom half) at the beam energy of 
63 MeV/nucleon,  
in the rest frame of the projectile, with $s$-wave (solid line) and
$d$-wave (dot-dashed line) configurations for the $^{11}$Be ground state.
Both calculations are done in FRDWBA and are normalized to the data peaks.
The data are taken from \protect\cite{mex}.
} 
\label{figg}
\end{figure}

In Fig. 8, we present the PMD of the $^{10}$Be fragment emitted in the
breakup of $^{11}$Be on U and Ta targets at 63 MeV/nucleon beam 
energy. Calculations performed within the FRDWBA and  
the AD models using both the $s$-wave and $d$-wave configurations 
are presented in this figure. 
The calculated cross sections are normalized to match 
the peak of the data points (which are given in 
arbitrary units) \cite{mex}, the normalization constant being the same 
for both cases. With the $s$-wave configuration the full width at
half maximum (FWHM) for the U and Ta targets are 44 MeV/c and 43 MeV/c,
respectively in both the FRDWBA and the AD cases. These agree well with
the averaged experimental value of 43.6$\pm$1.1 MeV/c \cite{mex}
and also with those calculated in \cite{pb1}. The very narrow widths
of the parallel momentum distributions signal the presence of a neutron
halo structure in $^{11}$Be.
It may be noted that the PMD calculated with a pure $d$ -- wave configuration
is too wide in width and grossly overestimates the experimental FWHM. 
 
In Fig. 9, we present the PMD (calculated within the FRDWBA formalism) 
of the $^{18}$C fragment in the breakup of $^{19}$C on a Ta target at the
beam energy of 88 MeV/nucleon.
\begin{figure}[ht]
\begin{center}
\mbox{\epsfig{file=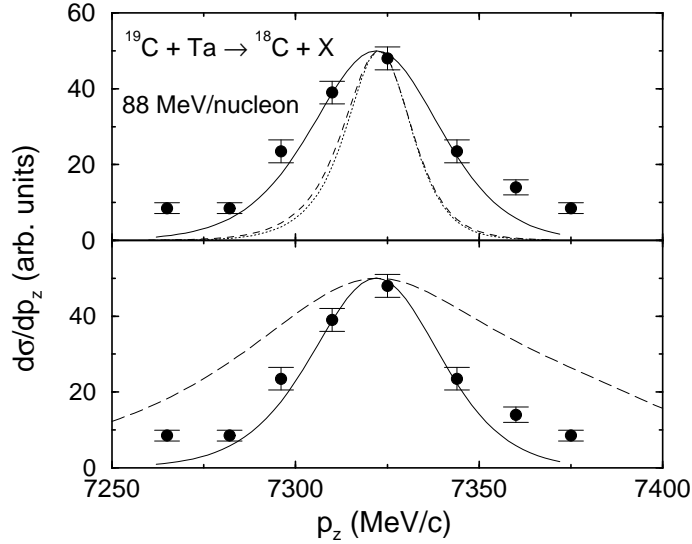,width=.75\textwidth}}
\end{center}
\caption[]{FRDWBA results for the parallel momentum distribution
of $^{18}$C in the breakup of $^{19}$C on Ta target at the beam energy
of 88 MeV/nucleon. The top half shows the results obtained with the
configuration
[$^{18}$C $(0^+)\otimes 1s_{1/2}\nu$] and single particle
wave function for the ground state of $^{19}$C with  
one-neutron separation energies of 530 keV (solid line), 160 keV (dashed line).
The dotted line shows the result obtained with a DCP wave function 
with a  one-neutron separation energy of 160 keV.
The bottom half shows the result obtained with the configurations
[$^{18}$C$(0^+)\otimes 1s_{1/2}\nu$] (solid) and
[$^{18}$C$(0^+)\otimes 0d_{5/2}\nu$] (dashed),  with the same value of the 
one-neutron separation energy (530 keV). The data have been
taken from \protect\cite{baz}.} 
\label{fig:figk}
\end{figure} 
We have normalized the peaks of the calculated PMDs to that of the data
(given in arbitrary units) \cite{baz} (this also involves coinciding the
position of maxima of the calculated and experimental PMDs). As can be
seen from the upper part of this figure, the experimental data clearly
favour $S_{n-^{18}C}$ = 0.53 MeV with the $s$ -- wave n-$^{18}$C relative
motion in the ground state of $^{19}$C (solid line).  The results obtained
with the $s$ -- wave configuration  within the simple potential
(dashed line) and DCP (dotted line) models (with the same value of 
$S_{n-^{18}C}=0.16~{\rm MeV}$) are similar to each other.

In the lower part of Fig. 9, we have shown the results obtained
with the $d$ -- wave relative motion (dashed line) for this system
(with $S_{n-^{18}C}$ = 0.53 MeV) and have compared it with that
obtained with a $s$ -- wave relative motion (solid line)
with the same value of the binding energy. As can be seen,
the FWHM of the experimental PMD is grossly overestimated by the $d$ -- wave
configuration.  The calculated FWHM with the $s$ -- state configuration
(with $S_{n-^{18}C}$ = 530 keV) is 40 MeV/c, which is in excellent agreement
with the experimental value of 41$\pm$3 MeV/c \cite{baz}. Thus these data
favour a configuration [$^{18}$C($0^+)\otimes 1s_{1/2}\nu$], with a one-neutron
separation energy of 0.530 MeV for the ground state of $^{19}$C. 
These results are in agreement with those of Ref. \cite{pb3}.
The narrow width of the PMD provides support to the presence of a 
one-neutron halo structure in $^{19}$C. 

We next consider the breakup of $^{15}$C which has a relatively larger
value for the one-neutron separation energy (1.2181 MeV)
and a ground state spin-parity of $1/2^+$ \cite{baz}. We
consider two configurations: a $1s_{1/2}$ 
neutron coupled to a $^{14}$C $(0^+)$ core and a $0d_{5/2}$ neutron 
coupled to a $^{14}$C $(0^+)$ core. One could have also considered a    
$^{14}$C $(2^+)$ core and $0d_{5/2}$ neutron coupling to get a $1/2^+$ ground
state for $^{15}$C, but it would raise the one-neutron separation 
energy to about 7.01 MeV, which is highly unfavourable for the formation
of a halo. We, therefore, do not consider this configuration in our study.
 
\begin{figure}[ht]
\begin{center}
\mbox{\epsfig{file=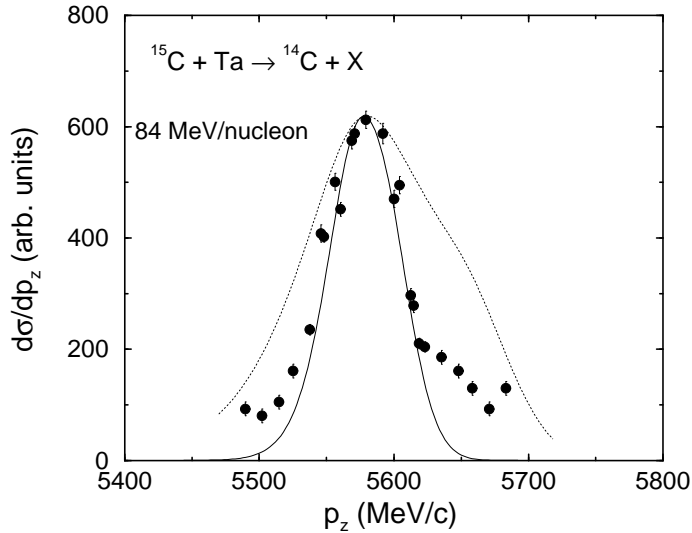,width=.75\textwidth}}
\end{center}
\caption{Parallel momentum distributions of $^{14}$C in the breakup of 
$^{15}$C on Ta at 84 MeV/nucleon. The solid line and dotted lines show the
results obtained with the configurations  
[$^{14}$C $(0^+)\otimes 1s_{1/2}\nu$] 
and [$^{14}$C $(0^+)\otimes 0d_{5/2}\nu$], respectively for the ground state of
the projectile. The data have been taken from \protect\cite{baz}.} 
\label{fig:figm}
\end{figure}

In Fig. 10, we present the results of our calculations for the PMD of
the $^{14}$C fragment in the breakup of $^{15}$C on a Ta target
at the beam energy of 84 MeV/nucleon. The experimental data are taken
from \cite{baz}. The $s$ -- state configuration for the ground state
of $^{15}$C gives a FWHM of 62 MeV/c, while with the $d$ -- state
configuration it comes out to be 140 MeV/c. Therefore, the experimental
value for the FWHM (67$\pm$1 MeV/c) \cite{baz} favours the former 
configuration. Thus, our results provide support to the existence 
of a halo structure in $^{15}$C. This nucleus provides an example of  
the one-neutron halo system with the largest one-neutron separation energy, 
known so far. 

$^{17}$C has a lower one-neutron separation energy (729 keV) as compared
to that of $^{15}$C. It would be interesting to see if it also has a 
halo structure, which seems probable if one considers only the binding
energies. The quoted ground state spin-parities for this nucleus are
$1/2^+, 3/2^+$ and $5/2^+$ \cite{rid97}. RMF calculations \cite{rmf}
predict it to have a value of $3/2^+$. We consider four possible ground
state configurations for this nucleus and calculate the parallel momentum
distributions of the $^{16}$C fragment in the breakup of $^{17}$C on a Ta
target at 84 MeV/nucleon beam energy within our FRDWBA formalism. 
The FWHMs of the PMD obtained with different configurations are listed
in Table 1.

\begin{table}
\begin{center}
\caption[T3]{FWHMs from the parallel momentum distribution of $^{16}$C for
different ground state configurations of $^{17}$C and one-neutron
separation energies ($\epsilon$) in the breakup of $^{17}$C on Ta
at 84 MeV/nucleon beam energy.}
\vspace{1.1cm}
\begin{tabular}{|c|c|c|}
\hline
Projectile & $\epsilon$ & $ FWHM $   \\
configuration& (\footnotesize{MeV})& (\footnotesize{MeV/c})  
 \\ \hline
$^{16}$C $(0^+)~~\otimes~~ 1s_{1/2}\nu$ & 0.729 & 51 \\
$^{16}$C $(0^+)~~\otimes~~ 0d_{5/2}\nu$ & 0.729 & 114 \\
$^{16}$C $(2^+)~~\otimes~~ 1s_{1/2}\nu$ & 2.5 &  82\\
$^{16}$C $(2^+)~~\otimes~~ 0d_{5/2}\nu$ & 2.5 & 185 \\ \hline
\end{tabular}
\end{center}
\end{table}

It is evident from this table that the $s$ -- state configurations predict
a narrow width for the PMD, providing support to the existence
of a halo structure in this nucleus. The experimental data \cite{baz}
for the breakup of $^{17}$C on a light target (Be) at 84 MeV/nucleon
give a FWHM of 145$\pm$5 MeV/c. Since the width of the PMD 
is mostly unaffected by the
reaction mechanism \cite{pb1}, it is quite likely that the experimental
FWHM will be the same also for the breakup of this nucleus on a 
heavier target. Therefore, the results shown in  
Table 1 seem to provide support to a $d$ -- wave configuration for the
ground state of $^{17}$C \cite{ang99}. Hence, the existence of a one-neutron
halo structure is quite improbable in 
$^{17}$C (see also Refs. {\cite{shyam_mad,udp}). 

In summary, from our pure Coulomb breakup studies we can say that 
the nuclei $^{11}$Be, $^{19}$C and $^{15}$C have a one-neutron halo
structure in their ground states. However, for $^{17}$C such a structure
appears to be less likely.  The calculated cross sections are selective
about the ground state wave function of the projectile. At the same
time there is a clear indication of the fact that only pure Coulomb
or pure nuclear breakup calculations may not be sufficient to describe
the details of the halo breakup data. Consideration of both modes of
breakup as well as of their interference terms is necessary to describe
the data properly.  

\subsection{Full breakup amplitude}

The full breakup amplitude that includes consistently both Coulomb 
and nuclear interactions between the projectile fragments and the target
nucleus to all orders has been developed in Refs.~\cite{raj02,raj03}.
We perform a Taylor series expansion of the distorted waves
of particles $b$ and $c$ about ${\bf r}_i$ and write
\begin{eqnarray}
\chi^{(-)}_b(\bq,{\bf r}) & = & e^{-i\alpha{\bf K}_b.\ro}
                           \chi^{(-)}_b(\bq,\ri), \\
\chi^{(-)}_c(\cq,{\bf r}_c) & = & e^{i\gamma{\bf K}_c.\ro}
                           \chi^{(-)}_c(\cq,\delta\ri). 
\end{eqnarray}
Employing the LMA \cite{shyam85,braun74a}, the magnitudes
of momenta ${\bf K}_j$ are taken as
\begin{eqnarray}
K_j(R)  = \sqrt {(2m_j/ \hbar^2)[E_j- V_j(R)]}, 
\end{eqnarray}
where $m_j$ ($j=b,c$) is the reduced mass of the $j-t$ system,
$E_j$ is the energy of particle $j$ relative to the target in the
center of mass (c.m.) system, and $V_j(R)$ is the potential between $j$ and
$t$ at a distance $R$. Substituting Eqs. (72) and (73) in Eq. (50),
and introducing the partial wave expansion of the distorted waves and
carrying out the angular momentum algebra, one gets 
\begin{eqnarray}
{\hat \ell} \beta_{\ell m} &=&
{(4\pi)^{3} \over {k_a k_b k_c\delta }} i^{-\ell} Y^\ell_{m_\ell}({\hat {\bf Q}})
Z_\ell (Q) \sum_{L_aL_bL_c} (i)^{L_a-L_b-L_c} {\hat L}_b{\hat L}_c
\nonumber \\
& \times & {\cal Y}^{L_b}_{L_c}({\hat k}_b,{\hat k}_c)
 \langle L_b 0 L_c 0| L_a 0 \rangle
 {\cal R}_{L_b,L_c,L_a}(k_a,k_b,k_a), 
\end{eqnarray}
where
\begin{eqnarray}
{\cal Y}^{L_b}_{L_c}({\hat k}_b,{\hat k}_c) & = &
\sum_M (-)^M\langle L_b M L_c -M|L_a 0 \rangle
Y^{L_b}_{M}({\hat {k}}_b)Y^{L_c*}_{M}({\hat {k}}_c), \\
Z_\ell (Q) & = & \int_0^{\infty} r_1^2 dr_1 j_{\ell}(Qr_1) u_\ell(r_1)V_{bc}(r_1),
 \\
{\cal R}_{L_b,L_c,L_a} & = & \int_0^{\infty}
{{dr_i}\over {r_i}} f_{L_a}(k_a,r_i)
f_{L_b}(k_b,r_i) f_{L_c}(k_c,\delta r_i). 
\end{eqnarray}
In Eq.~(75), Q is the magnitude of the vector
${\bf Q} = \gamma {\bf K}_c - \alpha {\bf K}_b$. Functions $f$
appearing in the radial integral ${\cal R}_{L_b,L_c,L_a}(k_a,k_b,k_a)$
are the radial parts of the distorted wave functions $\chi$'s. These are
obtained by solving the Schr\"odinger equation with appropriate optical
potential which include both Coulomb and nuclear terms.
The slowly converging radial integral ${\cal R}_{L_b,L_c,L_a}$ are
effectively handled by using the complex plane method~\cite{vin70,thesis}.

This theory can be used to calculate breakup of both neutron
and proton halo nuclei. Generally, the maximum value of the partial
waves $L_a,L_b,L_c$ must be very large in order to ensure the convergence
of the partial wave summations in Eq.~(75). However, for
the case of
one-neutron halo nuclei, one can make use of the following method
to include summations over infinite number of partial
waves. We write $\beta_{\ell m}$ as
\begin{eqnarray}
\beta_{\ell m} & = & \sum_{L_i = 0}^{L_{i}^{max}} {\hat \beta}_{\ell m} (L_i)
             + \sum_{L_i = L_{i}^{max}}^{\infty}
                    {\hat \beta}_{\ell m}(L_i), 
\end{eqnarray}
where ${\hat \beta}$ is defined in the same way as Eq.~(75)
except for
the summation sign and $L_i$ corresponds to $L_a$, $L_b$, and $L_c$. If
the value of $L_i^{max}$ is chosen to be appropriately large, the
contribution of the nuclear field to the second term of Eq.~(79)
can be neglected and we can write
\begin{eqnarray}
\sum_{L_i = L_{i}^{max}}^{\infty}{\hat \beta}_{\ell m}(L_i)  \approx
            \sum_{L_i = 0}^{\infty}{\hat \beta}_{\ell m}^{Coul}(L_i) -
             \sum_{L_i = 0}^{L_{i}^{max}}{\hat \beta}_{\ell m}^{Coul} (L_i),
\end{eqnarray}
where the first term on the right hand side, is the pure Coulomb
breakup amplitude which for the case where one of the outgoing fragments
is uncharged, can be expressed analytically in terms of the
bremsstrahlung integral (see, e.g., Ref. \cite{raj00}). Therefore, only
two terms, with reasonable upper limits, are required to be evaluated
by the partial wave expansion in Eq.~(79).

In the numerical applications for the breakup of $^{11}$Be, the structure
function $Z_\ell$ has been calculated with the $s$-wave configuration for
the $^{11}$Be ground state. The neutron-target optical potentials 
were extracted from the global set of Bechhetti-Greenlees
(see, e.g,~\cite{per76}), while those for the $^{10}$Be-target system
were taken from (\cite{per76,beu98}). Following~\cite{typ01},
we have used the sum of these two potentials for the
$^{11}$Be-target channel (see, \cite{raj03} for more details).
We found that values of $L_i^{max}$ of 500 for Au, Ta, U, Pb and Ti
targets and 150 for Be and C targets provided very good convergence
of the corresponding partial wave expansion series. The local momentum
wave vectors are evaluated at a distance, $R$ = 10 fm,
in all the cases with their directions being the same as that of
asymptotic momenta.

In Fig.\ 11, we show our results for the  neutron angular
distributions ($d\sigma/d\Omega_n$) for the breakup of $^{11}$Be on
Au, Ti and Be targets at 41 MeV/nucleon. The neutron
energy has been integrated from 26 MeV to 80 MeV, and the core scattering
angle in the lab system ($\theta_b$) has been integrated from $0^{\circ}$
to $30^{\circ}$ for the Au target case and from $0^{\circ}$
to $20^{\circ}$ for Ti and Be target cases. The dotted and dashed lines
represent the pure Coulomb and nuclear contributions, respectively while
their coherent sums are shown by the solid lines. The plus signs
and the inverted solid triangles represent the magnitudes of the
positive and negative interference terms, respectively.
Our calculations are in good
agreement with the experimental data~\cite{ann94}
(shown by solid circles) for all the three targets.

For the Be target, $d\sigma/d\Omega_n$
is governed solely by the nuclear breakup effects at all the angles.
The pure Coulomb breakup contributions are down by at least an
order of magnitude at the forward angles and by 2-3 orders of
magnitude at the backward angles. The CNI terms are also small
in this case.

On the other hand, for Ti and Au targets the Coulomb
terms are dominant at the forward angles while the nuclear breakup
effects are important at larger angles. This is to be expected
for high-Z targets, as the strong Coulomb field causes the fragile halo system
to breakup at large distances (and hence large impact parameters), leading
to the predominance of Coulomb breakup at forward angles. The nuclear breakup
assumes importance when the breakup occurs near the target nucleus,
consequently leading to large scattering angles.
\begin{figure}[h]
\begin{center}
\mbox{\epsfig{file=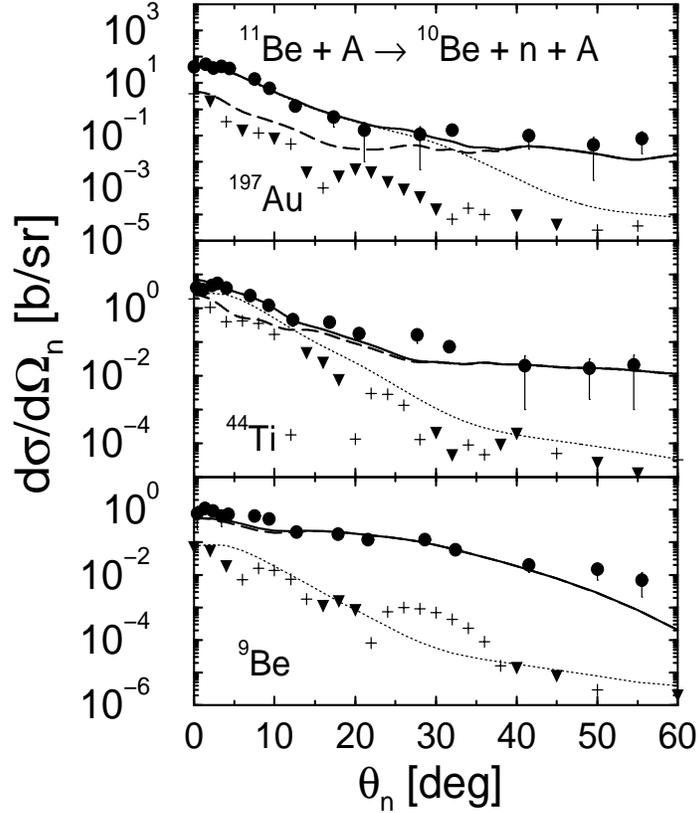,height=11cm,width=0.75\textwidth}}
\end{center}
\vskip .1in
\caption {
Neutron angular distribution for the breakup reaction $^{11}$Be + A
$\to$ $^{10}$Be + n + A at the beam energy of 41 MeV/nucleon.
The dotted and dashed
lines represent the pure Coulomb and nuclear contributions, respectively
while their coherent sums are shown by the solid lines. The plus signs
and the inverted solid triangles represent the magnitudes of the
positive and negative interference terms, respectively.
The data are taken from~\protect\cite{ann94}.
}
\end{figure}

Magnitudes of the CNI terms vary with angle; for many
forward angles they almost coincide with those
of the nuclear breakup while at the backward angles they are closer to the
pure Coulomb breakup contributions. Signs of these terms also change
with the neutron angle; a feature common to all the three targets.
It is clear that the interference terms are not negligible for
Ti and Au targets at the forward angles. For $\theta_n$
$\leq$ 10$^\circ$, the magnitudes of the CNI contributions are similar
to those of the pure nuclear terms, leading to a better description of the
data in this region.
\begin{figure}[ht]
\begin{center}
\mbox{\epsfig{file=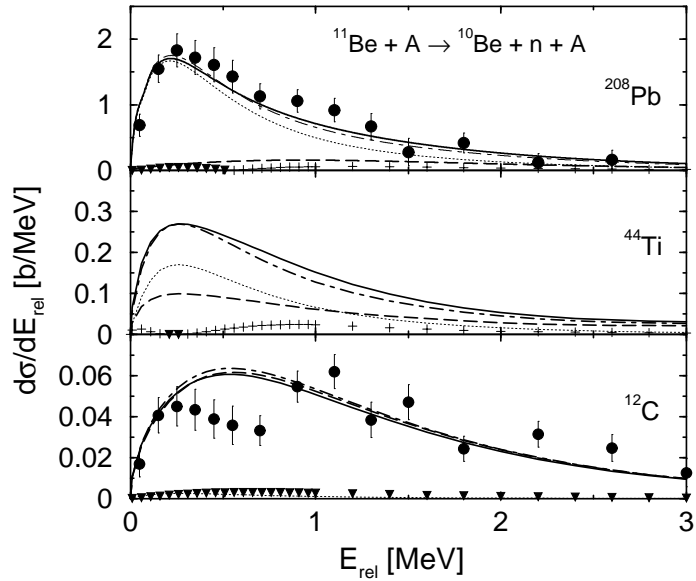,width=0.75\textwidth}}
\end{center}
\caption {
The differential cross section as a function of the relative energy of the
fragments (neutron and $^{10}$Be) in the breakup reaction of $^{11}$Be
on $^{208}$Pb, $^{44}$Ti and $^{12}$C targets at 72 MeV/nucleon.
The dotted and dashed
lines represent the pure Coulomb and nuclear breakup
contributions, respectively while their coherent and incoherent sums are
shown by the solid and dot-dashed lines, respectively.
The plus signs and the inverted triangles represent the magnitudes of the
positive and negative interference terms, respectively.
The data are taken from~\protect\cite{nak94}.
}
\end{figure}

The relative energy spectrum of the fragments (neutron and $^{10}$Be)
emitted in the breakup of $^{11}$Be on
$^{208}$Pb (top panel), $^{44}$Ti (middle panel) and
$^{12}$C (bottom panel) targets at the beam energy of 72 MeV/nucleon
is shown in Fig. 12.
In these calculations the integration over the projectile c.m. angle
($\theta_{n{^{10}{\rm Be}}-{\rm Pb}}$) has been done in the range of
$0^{\circ}$--$40^{\circ}$, mainly to include the effects of
nuclear breakup coming from small impact parameters. The relative
angle between the fragments ($\theta_{n-{^{10}{\rm Be}}}$)
has been integrated from $0^{\circ}$ to $180^{\circ}$.
The dotted and dashed
lines represent the pure Coulomb and nuclear breakup
contributions, respectively while their coherent and incoherent sums are
shown by the solid and dot-dashed lines, respectively.
The plus signs and the inverted triangles represent the magnitudes of the
positive and negative interference terms, respectively.

In case of breakup on a heavy target ($^{208}$Pb) [Fig. (12)(top panel)]
the pure Coulomb contributions dominate the cross sections around
the peak value, while at larger relative energies nuclear breakup gains importance.
This is attributed to the different energy dependence of the two
contributions~\cite{typ01}.
The nuclear breakup occurs when the projectile and the
target nuclei are close to each other. Its magnitude, which is determined
mostly by the geometrical conditions, has a weak dependence
on the relative energy of the outgoing fragments beyond a certain
minimum value. In contrast, the Coulomb breakup contribution has a
long range and it shows a strong energy dependence.
The number of virtual photons increases for small excitation energies
and hence the cross sections rise sharply at low excitation energies.
After a certain value of this energy the cross sections decrease due to
setting in of the adiabatic cut-off.  The coherent sum of the Coulomb and
nuclear contributions provides a good overall description of the experimental
data. The nuclear and the CNI terms are necessary to explain the data at
larger relative energies.

In the middle panel of Fig.~(12), we show the relative energy
of the fragments in the breakup of $^{11}$Be on a medium mass target
($^{44}$Ti). At low relative energies the pure Coulomb contributions
are slightly higher than the pure nuclear ones, while at higher relative
energies it is the nuclear part which dominates. Apart from the very low
relative energy region the CNI terms play an important role, which is
clearly borne out by the difference in the coherent (solid) and incoherent
(dot-dashed) sums of the pure Coulomb and pure nuclear contributions.

The relative energy spectra for the breakup on a light target ($^{12}$C)
is shown in the bottom panel of Fig. (12). In this case we have used the
same optical potential for the $^{10}$Be-$^{12}$C system as in the
$^{10}$Be-$^{9}$Be case, which we had used earlier in calculating the
neutron angular distribution in Ref. \cite{raj02}. The total cross section
in this case is normalized to the experimental cross section (found by
integrating the area under the data points) and the same normalization
constant is used for all the cross sections in this case.
The breakup is clearly seen to be
nuclear dominated at all relative energies, and the pure Coulomb and CNI terms
have very little contributions.

The parallel momentum distributions (PMDs) of the $^{10}$Be fragment
in the breakup of $^{11}$Be
on U and Ta targets, at 63 MeV/nucleon beam energy are presented in the rest
frame of the projectile, in Fig. 13.
The dotted and dashed lines show the contributions of the pure Coulomb
and nuclear breakups, respectively, while their coherent sums are 
represented by solid lines. The coherent sum is normalized to the peak of the data,
which are given in arbitrary units, and the same normalization factor
has been used for the pure Coulomb and pure nuclear contributions.
\begin{figure}[ht]
\begin{center}
\mbox{\epsfig{file=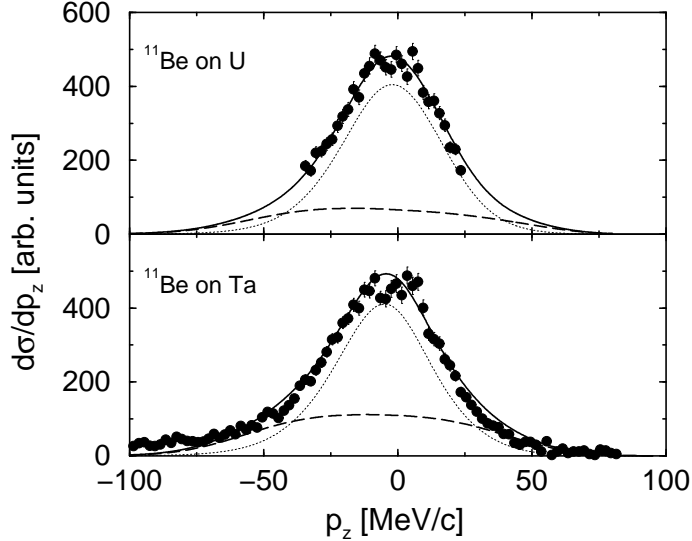,width=0.75\textwidth}}
\end{center}
\caption {
The parallel momentum distribution of the core
 in the breakup of $^{11}$Be
on U and Ta targets, at 63 MeV/nucleon beam energy, in the rest
frame of the projectile. The dotted and dashed
lines represent the pure Coulomb and nuclear breakup
contributions, respectively while their coherent sums are
shown by solid lines. The data are taken from~\protect\cite{kelly95_5}.
}
\end{figure}

The above results make it clear that the Coulomb-nuclear interference
terms are both energy and angle dependent.
They are almost of the same magnitude as the nuclear breakup
contributions in the neutron angular distributions on heavy and medium 
mass targets. This leads to a difference in the coherent and incoherent
sums of the Coulomb and nuclear terms, more so at forward angles.
In the parallel momentum distribution of the $^{10}$Be fragment in the breakup
reaction of $^{11}$Be the region around the peak of the distribution,
which gets substantial contributions from forward scattered fragments,
is Coulomb dominated, while in the wings of the distribution, where
contributions come from fragments scattered at large angles, the nuclear
breakup contributions dominate.

The relative energy spectra of the fragments
(neutron and $^{10}$Be) are largely nuclear dominated for light targets.
However, to explain satisfactorily the data on heavier targets  
one requires both the nuclear and the CNI terms particularly at
higher relative energies. In case of breakup on a medium mass target,
the total pure Coulomb and pure nuclear contributions were nearly equal
in magnitude. Thus in many sophisticated experiments planned in the
future one has to look into the role played by the CNI terms in
analyzing the experimental data.

%%%%%%%%%%%%%%%%%%%% E3 %%%%%%%%%%%%%%%%%%%%%%%%%%%%%%%%%%%%%%%%%
%%%%%%%%%%%%%%%%%%%% B4 %%%%%%%%%%%%%%%%%%%%%%%%%%%%%%%%%%%%%%%%%
\section{Postacceleration effects in the Coulomb breakup of
neutron halo nuclei}

As shown section 3.1, an important advantage of the post form DWBA
theory of breakup reactions is that it can be solved analytically
for the case of the breakup of the neutron halo nuclei with the entrance
and outgoing channels  involving only the Coulomb distortions
\cite{btr72,shyam92}. It constitutes an ideal ``theoretical laboratory"
to investigate the physics of the breakup reactions, its certain limiting
cases, and its relation to other models like the semiclassical approximation.
Particularly, the effect of postacceleration can be studied in a unique
way within this approach.

Postacceleration refers to the situation where the core $c$ has
a larger final state energy than what one gets from sharing the kinetic
energy among the fragments according to their mass ratio. This effect
arises in  a purely classical picture \cite{BaurBK95} of the breakup process.
The nucleus $a=(c+n)$ moves up the Coulomb potential, loosing the
appropriate amount of kinetic energy. At an assumed ``breakup point'',
this kinetic energy (minus the binding energy) is supposed to be shared
among the fragments according to their mass ratio (assuming that the
velocities of $c$ and $n$ are equal). Running down the Coulomb barrier, the
charged particle $c$ alone (and not the neutron) gains back the Coulomb 
energy, resulting in its postacceleration. Of course this picture is based
on the purely classical interpretation of this process, and will be
modified in a quantal treatment, where such a ``breakup point'' does
not exist.  Postacceleration is clearly observed in the low energy deuteron
breakup, both in the theoretical calculations
as well as in the corresponding experiments (see, e.g., \cite{baur84,BaurT76}).
However, in the description of the Coulomb dissociation of halo nuclei at
high beam energies within this theory \cite{raj00,shyam92,shyam93},
the postacceleration effects become negligibly small. We shall investigate
this point further for the $^{11}$Be and  $^{19}$C Coulomb dissociation
experiments \cite{nak94,nak99}. 
On the other hand, in the semiclassical Coulomb excitation theory
the higher order effects have been found \cite{TypelB01} to be small,
for both zero range as well as finite range wave functions of the $c+n$ system. 

It was recently noticed \cite{baur03}
that in the limit of Coulomb parameter $\eta_a \ll 1$ (i.e. in the Born
approximation), the pure Coulomb post form DWBA [Eq. (53)] leads to
results which are same as those obtained in a semiclassical model 
{\cite{bht01}}.
This agreement is also valid for arbitrary values of $\eta_a$ and
$\eta_c$, provided the beam energies are high as compared 
to the relative energy ($E_{cn}$) of fragments $c$ and $n$ in the ground
state of the projectile. The first order approximation to
the amplitude given by Eq.~(53), can be written as \cite{bht01,pb02,baur03} 
\begin{eqnarray}
{\hat \ell}\beta^{\textrm {first order}} _{\ell m} & = & 4\pi Z_{\ell m} 
f_{\textrm{coul}} e^{-\frac{\pi}{2}\xi}\nonumber \\ \times
&&\left[e^{-i\phi}\frac{1}{k_a^2 - \left[\vec k_c + \delta\vec k_n \right]^2}
+ e^{i\phi}\frac{m_c}{m_a}\frac{1}{\left[k_c^2 - \left(\delta\vec k_n -
\vec k_a\right)^2 \right]} \right],
\end{eqnarray} 
where the relative phase $\phi = \sigma(\eta_c) - \sigma(\eta_a) -
\sigma(\xi) - \xi/[2\log|D(0)|]$, with $\sigma(\eta)$ being the usual
Coulomb phase shifts, and $\xi = \eta_c - \eta_a$ and $D(0)$ as defined in
Eq. (70). In Eq. (81), we have 
defined $f_{\textrm {coul}}= 2\eta_a k_a/k^2$. This term is very similar 
to the Born approximation (BA) result given in \cite{BaurHT01};
in the limit $\xi\rightarrow 0$ it actually coincides with the BA expression. 
This equation can be used to investigate the role of higher order effects
(which includes postacceleration). It may be noted that the derivation
of Eq.~(81) makes use only of the condition 
$-D(0) \gg 1$ which is met for beam energies large as compared
to the binding energy.
\begin{figure}
\begin{center}
\mbox{\epsfig{file=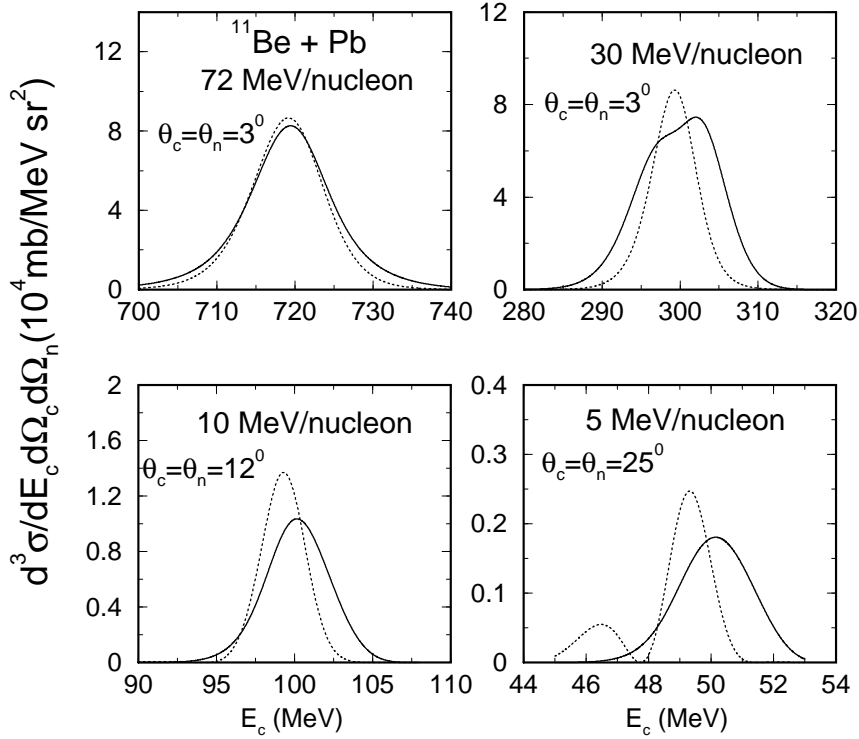,height=10.0cm}}
\end{center}
\caption {
Triple differential cross section as a function of the energy of 
$^{10}$Be core for the reaction $^{11}$Be + Pb $\rightarrow$ n + $^{10}$Be
+ Pb at the beam energies of 72 MeV/nucleon, 30 MeV/nucleon, 10 MeV/nucleon
and 5 MeV/nucleon. The results of the finite range DWBA and first-order
theory are shown by solid and dotted lines respectively.
}
%\label{fig:figa}
\end{figure}
\noindent

We now investigate postacceleration effects in the breakup of
the one-neutron halo nuclei $^{11}$Be
and $^{19}$C.  We take a heavy target of atomic number Z=82.  
In the following all the higher order results correspond to calculations 
performed within the finite range post form DWBA theory as discussed in
section 3.1. The structure inputs were also the same as those given
their. The first order results have been obtained by using Eq.~(81). 

In Fig.~(14), we present calculations for the triple differential
cross sections for the breakup reaction
$^{11}$Be + Pb $\rightarrow$ n + $^{10}$Be + Pb, as a function of the
energy of the $^{10}$Be core (E$_c$), for four beam energies
lying in the range of 5 MeV/nucleon - 72 MeV/nucleon. To see the
postacceleration in a clear way, it is very useful to study the cross-section
as a function of the core energy. The results obtained within the higher order
and the first order theories are shown by solid and dotted lines, respectively.

It can be seen from this figure that while for lower beam energies, the
higher order and first-order results differ considerably from each other,
they are almost the same for the beam energy of 72 MeV/nucleon. In each case,
the first order cross sections peak at the energy of the core fragment which
corresponds to the beam velocity (this value of the core fragment energy
will be referred to as E$_{bv}$ in the following). In contrast to this,
the peaks of the higher order cross sections are shifted to
energies $>$ E$_{bv}$ for the three lower energies. Only for the
72 MeV/nucleon beam energy, does the higher order result peak at E$_{bv}$.
This shows very clearly that the finite range DWBA model exhibits
postacceleration for beam energies $\leq$ 30 MeV/nucleon, while this
effect is not present at 72 MeV/nucleon. Therefore, the higher order effects
are minimal for the Coulomb breakup of $^{11}$Be at the beam energies
$\geq$ 70 MeV. This result is in agreement with those obtained in
\cite{TypelB01,typ01}.

In Fig.\ (15), we compare the results of the first-order and the
finite range DWBA calculations for the relative energy spectrum of the
fragments emitted in the breakup reaction of $^{11}$Be on a $^{208}$Pb
target for the same four beam energies as shown in Fig.\ (14). These
cross sections have been obtained by integrating over all the allowed
values of the angles $\Omega_{c-n}$. In both the models, the integrations
over $\theta_{t-(c+n)}$, have been carried out between 1$^\circ$ to
grazing angle, in the upper two figures, and between 5$^\circ$ to grazing
angles, in the lower two figures. The integrations over $\phi_{t-(c+n)}$
angles have been done over all of its kinematically allowed values. The
dotted and solid lines represent the first-order and the higher order
results, respectively. 
\begin{figure}[here]
\begin{center}
\mbox{\epsfig{file=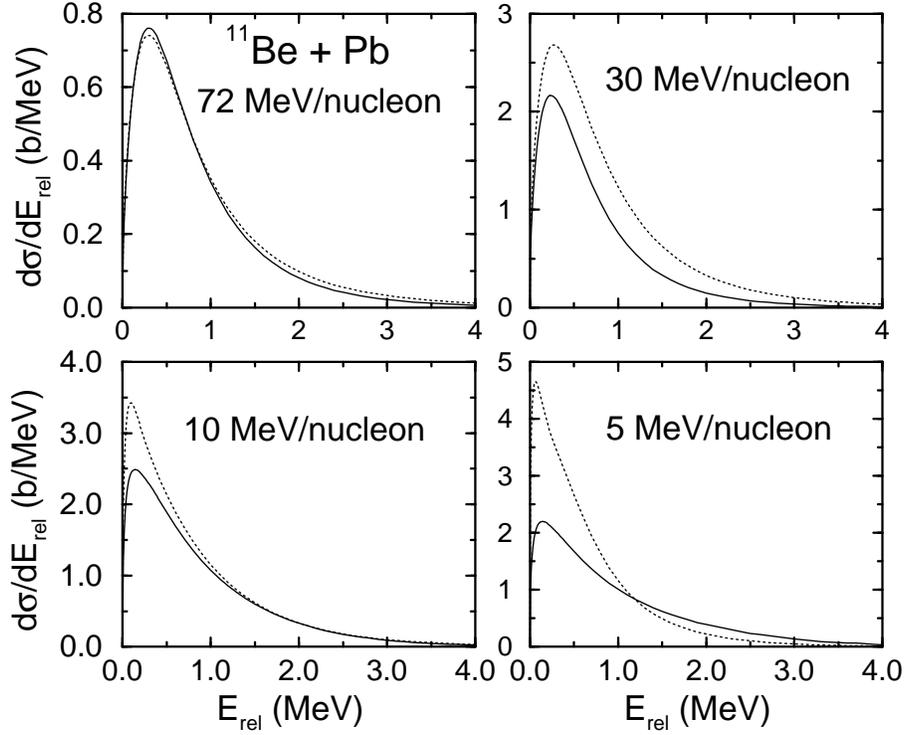,height=10.0cm}}
\end{center}
\caption {
The differential cross section as a function of the relative energy of the
fragments (neutron and $^{10}$Be) emitted in the $^{11}$Be induced
breakup reaction on a $^{208}$Pb target at the beam energies of
72 MeV/nucleon, 30 MeV/nucleon, 10 MeV/nucleon, and 5 MeV/nucleon.
The dotted and full lines represent the first-order and the finite range 
DWBA results, respectively.  
}
%\label{fig:figc}
\end{figure}
\noindent

We notice that while for the beam
energy of 72 MeV/nucleon, the higher order effects are
minimal, they are quite strong  
for the lower beam energies, being largest at the beam energy of
5 MeV/nucleon. This reinforces the point, already made 
in \cite{TypelB01,typ01}, that at the beam energy of 72 MeV/nucleon, the
higher order effects are quite small if both the first order and the higher
order terms are calculated within the same theory.

In Fig.\ (16), we show the same results as in Fig. (15) but for the  
$^{19}$C induced reaction on the $^{208}$Pb target for the beam 
energies of 67 MeV/nucleon, 30 MeV/nucleon and 10 MeV/nucleon. 
We see that in this case too the higher order effects are quite weak
for the beam energy 67 MeV/nucleon, but appreciable for the lower
beam energies.
\begin{figure}[here]
\begin{center}
\mbox{\epsfig{file=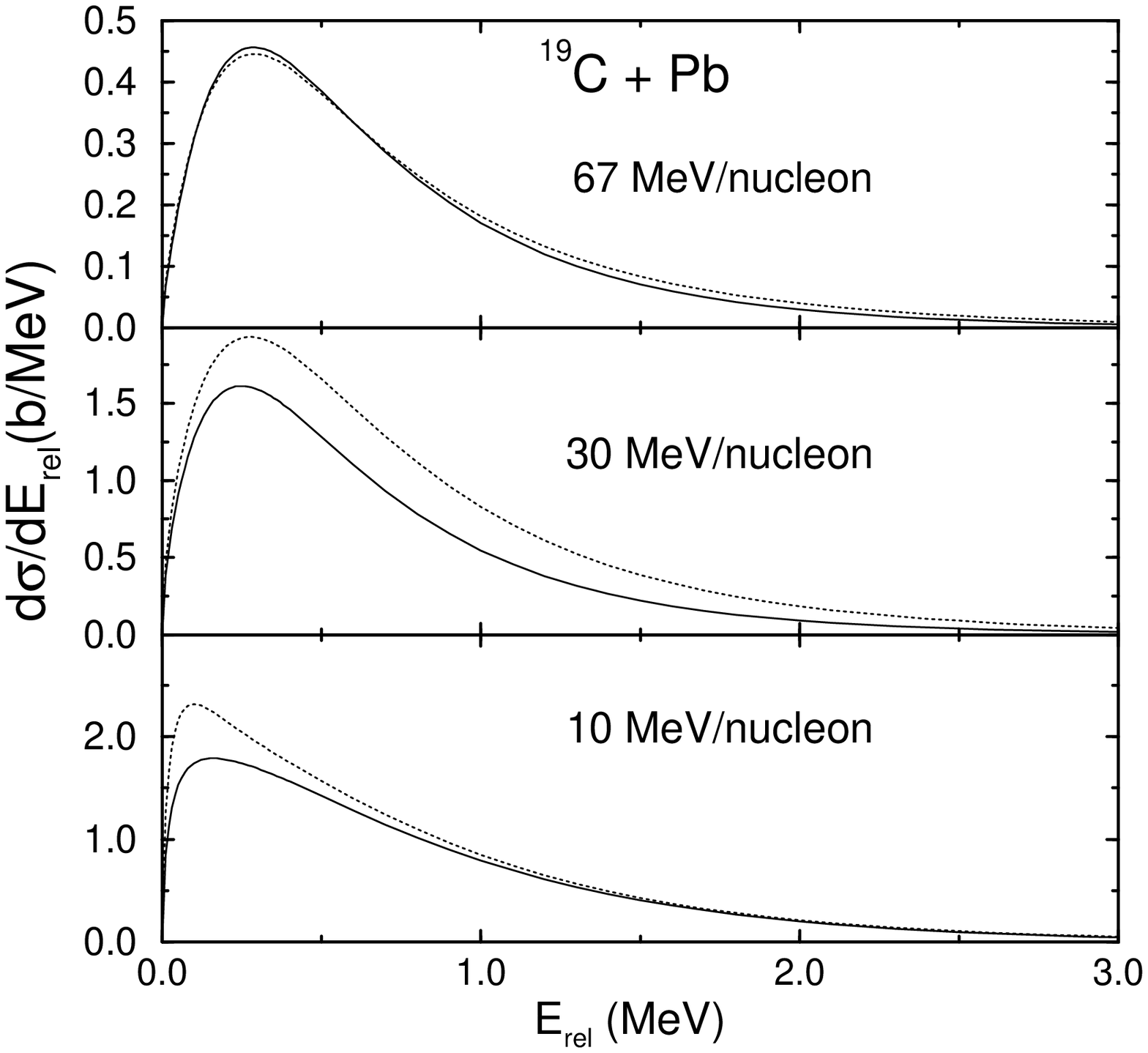,height=10.0cm}}
\end{center}
%\vskip .3in
\caption{
The differential cross section as a function of the relative energy of the
fragments (neutron and $^{18}$C) emitted in the $^{19}$C induced
breakup reaction on a $^{208}$Pb target at the beam energies of
67 MeV/nucleon, 30 MeV/nucleon, and 10 MeV/nucleon.
The dotted and full lines represent the first-order and the finite range 
DWBA results, respectively.  
}
%\label{fig:figd}
\end{figure}
\noindent
 
It may be noted that by comparing the result of the adiabatic model
of Coulomb breakup reactions which is conceptually different from 
ours, with that of the first order semiclassical perturbation
theory of the Coulomb excitation,
it has been concluded in \cite{tos00} that the higher order effects are
substantial for these reactions even at the beam energies $\sim$
70 MeV/nucleon. However, one should be careful in drawing definite
conclusions about the role of the higher order effects from such an
approach. For a reliable assessment of the contributions of the higher order
effects, it is essential that both the first order and the
higher order terms should be  calculated within the same theory, as has
been done in \cite{TypelB01,typ01,pb02}).

Thus, in the post form DWBA theory, the peaks in
the triple and double differential cross sections vs. core energy spectra,
are shifted to energies larger than those corresponding to the beam
velocity, at the incident energies $\leq$ 30 MeV/nucleon. Therefore,
postacceleration effects are important at these beam energies. On the
other hand, at the beam energy $\sim$ 70 MeV/nucleon, the corresponding
spectra peak at the beam velocity energies, which is consistent with
no postacceleration. In contrast to this, the first-order cross sections
always peak at the beam velocity energy, which is expected as the
postacceleration is a higher order effect. 

The higher order effects are also found to be quite important in the
relative energy spectrum of the fragments at beam energies $\leq$ 30 
MeV/nucleon, while they are insignificant at the beam energies $\sim$
70 MeV/nucleon. This suggests that the conclusions arrived at in Refs.
\cite{nak94,nak99}, where the data on the relative energy spectra of
the fragments taken in the breakup of $^{11}$Be and $^{19}$C at 
beam energies $\sim$ 70 MeV, have
been analyzed within the first order theory of the Coulomb excitation, 
may not be affected by the higher order effects. 
 
It should be noted that from an experimental point of view, the
postacceleration effects are not fully clarified
(see, e.g., \cite{nak94,Ieki93,Bush98}).
Finally, let us mention the recent work on the electromagnetic dissociation
of unstable neutron-rich oxygen isotopes \cite{Leistenschneider01}.
These authors  deduce photoneutron cross sections from their dissociation
measurements. If the neutrons are emitted in a slow evaporation process in
a later stage of the reaction, the question of postacceleration is not
there. On the other hand, for the light nuclei there is some direct
neutron emission component and the present kind of theoretical analysis
further  proves the validity of the semiclassical approach used
in \cite{Leistenschneider01}.

Postacceleration effects are also of importance for the use of Coulomb 
dissociation for the study of radiative capture reactions of
astrophysical interest. We expect that our present investigations will
shed light on  questions of  postacceleration and 
higher order effects in these cases also.

%%%%%%%%%%%%%%%%%%%% E4 %%%%%%%%%%%%%%%%%%%%%%%%%%%%%%%%%%%%%%%%%
%%%%%%%%%%%%%%%%%%%% B5 %%%%%%%%%%%%%%%%%%%%%%%%%%%%%%%%%%%%%%%%%
\section{Summary, Conclusions and Future Outlook}

Due to the potential use of the breakup reactions of neutron rich 
light exotic nuclei in extraction the information about the structure
of these nuclei, it is essential to have a full quantal
theory of these reactions which involves minimum number of parameters. 
To this end, we have developed a theory of one-neutron halo nuclei
within the framework of the post form distorted wave Born approximation.
Finite range effects of the core-halo interaction are
included in this theory which allows the full ground state wave function
of the projectile corresponding to any orbital angular momentum
structure to enter into this theory. For the one-nucleon halo breakup case,
it can treat the Coulomb and nuclear breakups as well as their interference
terms within a single framework on an equal footing. Since this theory uses
the post form scattering amplitude, the breakup contributions from the
entire continuum corresponding to all the multipoles and relative
orbital angular momenta of the core-halo system are included in it. This
theory can account for the postacceleration effects in a unique way.

Most of the breakup observables are sensitive to the ground state
configuration of the projectile. We find that for $^{11}$Be,
a $s$ -- wave configuration [$^{10}$Be$(0^+) \otimes 1s_{1/2}\nu$], with a 
spectroscopic factor of 0.74 for its ground state 
provides best agreement with the experimental data in all the cases. 

We have also performed calculations within an adiabatic model which makes
the approximation that the strongly excited core-valence particle
relative energies are small in the Coulomb breakup and also within
the approximation of Baur and Trautmann which equates the coordinates
of the core-target system with those of the projectile-target system.
Unlike the DWBA, the adiabatic model
does not use the weak coupling approximation to describe
the center of mass motion of the fragments with respect to the target.
For almost all the observables, there is a general agreement between
the DWBA and the adiabatic model results even in the absolute
magnitude. However, the approximation of Baur and Trautmann gives
results which are very different from those obtained within the
DWBA and the adiabatic model.

For the $^{19}$C case, the results for the PMD of $^{18}$C 
and the relative energy spectrum of the $n$ - $^{18}$C system
show that the most probable ground state configuration of $^{19}$C is
[$^{18}$C $(0^+)\otimes 1s_{1/2}\nu$] with a one-neutron separation
energy of 530 keV and a spectroscopic factor of 1. Furthermore, 
the most probable configuration for $^{15}$C is a $s$ -- wave valence
neutron coupled to the $^{14}$C core and that for $^{17}$C is a 
$d$ -- wave valence neutron coupled to the $^{16}$C core. Both the 
experimental and the calculated FWHM of the PMD for the $^{14}$C core
in the breakup of $^{15}$C are small and they agree well with each other.
This provides support to the existence of a one-neutron
halo structure in $^{15}$C. On the other hand, in the case of $^{17}$C the 
FWHM of the PMD for the $^{16}$C core is closer to that of
a stable isotope. Therefore the existence of a halo structure in
$^{17}$C appears to be unlikely. Interestingly the one-neutron
separation energies of $^{15}$C and $^{17}$C are 1.2181 and 0.729 MeV, 
respectively. So both the binding energy of the valence neutron as well
as its configuration with respect to the core together decide whether
a nucleus has halo structure or not. 

For medium mass and heavy target nuclei, the neutron angular distributions 
are dominated by the nuclear and the Coulomb breakup
terms at larger and smaller angles, respectively. Contributions
of the Coulomb-nuclear interference terms are non-negligible.
They can be as big in magnitude as the pure nuclear or the pure
Coulomb breakup and have negative or positive sign depending upon
the angle and energy of the outgoing fragments. For these targets,
the interference terms help in better description of trends of
the experimental data even at smaller angles. 

In the parallel momentum distribution of the $^{10}$Be fragment in the breakup
reaction of $^{11}$Be, the region around the peak of the distribution,
which gets substantial contributions from forward scattered fragments,
is Coulomb dominated, while in the wings of the distribution, where
contributions come from fragments scattered at large angles, the nuclear
breakup contributions dominate.  Similarly, the data on
relative energy spectra of the fragments (neutron and $^{10}$Be)
emitted in breakup of $^{11}$Be on a heavy target 
can not be described properly by considering only the pure Coulomb
breakup mechanism; inclusion of nuclear and Coulomb-nuclear interference
terms is necessary.

In future studies the full quantal theory of the one-neutron halo
breakup reactions should  be applied to describe the $(a,b\gamma)$ reaction.
The data for these reactions \cite{nav00,val00}
taken at the Michigan State University are yet to be described within a
full quantal reaction model. Since, the relevant partial cross sections
of the core fragments are essentially inclusive, both elastic and
inelastic breakup modes will contribute to them. Our theory can
be extended to calculate the latter mode in a straight forward
manner. In fact all the ingredients required for this extension
have already been calculated by us.

There is also a need to extend the theory to describe the halo
breakup at higher beam energies for which data have been taken
at GSI, Darmstadt. This can be achieved by introducing the
eikonal expansion of the distorted waves, instead of the partial
wave expansion as done in section 3.2.

Our theory should be used to analyze the breakup
data of $^{8}$B on a $^{58}$Ni target at the beam energy of 25.8 MeV. 
At this energy the
nuclear breakup effects are quite strong and the Coulomb-nuclear
interference terms should manifest themselves in an important way. It should
be noted that in the CDCC model analysis of these data, the continuum
states corresponding to much larger excitation energies and relative
orbital angular momenta were required to be included in order
to get a proper convergence. Our post form breakup theory
includes, by its very construction, contributions from the entire
continuum corresponding to all the excitation energies,
multipoles and relative orbital angular momenta of the core-valence
nucleon system. Therefore, a comparison of our calculations
with these data would be interesting also from the point of view of
checking and supplementing the corresponding CDCC results.

The study of the $^8$Li($\alpha$,$n$)$^{11}$B 
reaction is of crucial importance in determining the abundance of
$^{11}$B, since nuclides with A $\geq$ 12 pass through $^{11}$B on their
way to higher masses.
However, the $\alpha$-capture reaction $^8$Li($\alpha$,$n$)$^{11}$B is in 
competition 
with the $^8$Li($n$,$\gamma$)$^9$Li reaction which turns the reaction flow
back to lighter elements via $^9$Li($\beta^-$,$\nu$)$^9$Be($p$,$\alpha$)$^6$Li.
Thus the cross sections of these reactions are extremely 
important in predicting the yields of elements with A $\geq$ 12.
We would like to apply our theory to investigate
the breakup reaction $^9$Li $\to $ $^8$Li + $n$ on a heavy 
target like $^{208}$Pb.
It would be interesting to know how far the data taken recently at the
Michigan State University can be described solely by the Coulomb
breakup mechanism and what is the role of the nuclear breakup
effects in these data. This is important as the pure Coulomb breakup
cross sections for this reaction can be used to extract the cross
sections for the reverse reaction $^8$Li($n$,$\gamma$)$^9$Li.

The authors would like to express their thanks to P. Banerjee, G. Baur,
P. Danielewicz and H. Lenske for several useful discussions on
the present topic.

%%%%%%%%%%%%%%%%%%%% E5 %%%%%%%%%%%%%%%%%%%%%%%%%%%%%%%%%%%%%%%%%
%%%%%%%%%%%%%%%%%%%% B bibl %%%%%%%%%%%%%%%%%%%%%%%%%%%%%%%%%%%%%%%%%

%%%%%%%%%%%%%%%%%%%% E bibl %%%%%%%%%%%%%%%%%%%%%%%%%%%%%%%%%%%%%%%%%

\end{document}